\documentclass[prd,tightenlines,nofootinbib,amsfonts,amssymb,amsmath,color,11pt]{article}
\usepackage[utf8]{inputenc}
\usepackage[affil-it]{authblk}
\usepackage{amsmath}
\usepackage{amssymb}
\usepackage{hyperref}
\usepackage{geometry}
\usepackage{color}
\usepackage{graphicx}
\usepackage{subfig}

\newcommand{\dif}{\mathrm{d}}
\newcommand{\me}{\mathrm{e}}
\newcommand{\mi}{\mathrm{i}}

\newcommand{\de}{\partial}
\newcommand{\mpl}{M_{\rm Pl}}

\def\be{\begin{equation}} 
\def\ee{\end{equation}}
\def\bea{\begin{eqnarray}}
\def\eea{\end{eqnarray}}

\begin{document}

\title{\bf Distinctive signatures of space-time \\diffeomorphism breaking in EFT of inflation}
\author[1,2]{Nicola Bartolo\thanks{nicola.bartolo@pd.infn.it}}
\author[1,2]{Dario Cannone\thanks{dario.cannone@pd.infn.it}}
\author[3]{Angelo Ricciardone\thanks{angelo.ricciardone@uis.no}}
\author[4]{Gianmassimo Tasinato\thanks{g.tasinato@swansea.ac.uk}}

\affil[1]{\small Dipartimento di Fisica e Astronomia “G. Galilei”, \break Universit\`a degli Studi di Padova, via Marzolo 8, I-35131, Padova, Italy}
\affil[2]{INFN, Sezione di Padova, via Marzolo 8, I-35131, Padova, Italy}
\affil[3]{Faculty of Science and Technology, University of Stavanger, 4036, Stavanger, Norway}
\affil[4]{Department of Physics, Swansea University,
Swansea, SA2 8PP, U.K.}

\date{\today}
 \maketitle

 \abstract{
   The effective field theory of inflation is a powerful tool for obtaining model independent predictions common to large classes
of inflationary models. It requires only information about the symmetries broken during the inflationary era, and on the number
and nature of fields that drive inflation. In this paper, we consider the case for scenarios that  simultaneously break  time 
reparameterization and spatial diffeomorphisms during inflation. We examine how to analyse such systems using an effective field
theory approach, and we discuss several  observational consequences 
  for the statistics of scalar and tensor modes. For example, examining the three point functions, we show that this symmetry breaking pattern can lead to an enhanced amplitude  for  the squeezed  bispectra, and to a distinctive angular dependence between their three wavevectors.
We also discuss how our results indicate prospects for constraining the level of spatial diffeomorphism breaking during inflation.}
  
 \bigskip
\bigskip
 
 \section{Introduction}

 The most precise cosmological observations at our disposal \cite{Ade:2015lrj, Ade:2015ava}
  are compatible with simple scenarios of inflation based on a single field model, that lead to 
an almost scale invariant, gaussian spectrum of scalar adiabatic perturbations.
 On the other hand, both the wealth of more precise
  data that will become available in the next few years, and the fact that some anomalies seem to be 
  present in current observations (see~\cite{Ade:2015hxq} and, e.g., 
 \cite{Schwarz:2015cma}  for a review) 
 suggest to keep an eye open to
  different
   possibilities.

The effective field theory of inflation (EFTI) is  a powerful tool 
 for obtaining model independent  predictions for large classes of inflationary scenarios. It  requires  only    information
 about the symmetries broken during the inflationary era, and on the number and nature of fields that drive inflation. 
The advantage of
EFTI is that  one has not to commit on specific realizations in order to deduce  observable consequences.
   One writes
the most general set of operators
 that satisfies the symmetry requirements and connects the coefficients of such operators with observable quantities. 
 The EFTI has been successfully applied to cosmological models of single-field inflation, that break time reparameterization.
 
 Space-time diffeomorphisms correspond to the invariance of the theory under the gauge symmetry of General Relativity:
 \be \label{defdi}
 x^\mu \to x'^{ \mu}( x^\nu)\;,
 \ee
 for arbitrary functions $ x'^{ \mu}( x^\nu)$
 of the coordinates.  
 During inflation,  time-reparameterisation  invariance 
 \be t\to t+\xi(x^\mu)\;,\ee
  is broken. This is due to the existence of an `inflationary clock'  that 
 breaks time diffeomorphisms and controls how much time is left before inflation ends. In models
  of single field inflation,  there is a unique clock and only adiabatic modes can be generated on superhorizon scales. What is controlling the clock dynamics is what sources
  inflation, and at the same time causes  the spontaneous  breaking of de Sitter symmetry during 
   the 
  inflationary era.
  Studying the system at high energies, we can expect that gravity decouples. Time reparameterisation
      becomes a global symmetry, and its breaking gives  rise
  to a  massless Goldstone boson $\pi$. Its high energy  
   dynamics faithfully describes the dynamics of the fluctuations of  
    the inflationary clock.  This thanks to Goldstone boson equivalence theorems, originally  proven 
   in quantum field theory for gauge symmetries \cite{Cornwall:1974km}, and more recently  applied to the EFTI in \cite{Cheung:2007st,Baumann:2011su}.

While time reparameterization invariance is certainly spontaneously  broken by the
 source of inflation, it is interesting to explore  the 
possibility that {\it space diffeomorphisms} are also broken during the inflationary epoch. After all, we are ignorant 
 about what was really happening at the  high energy scales and early times  characterising inflation. 
  In this paper, we consider  cases where also the symmetry
 \be
 x^i \to x'^{i}(t,\,x^j)\;,
 \ee
 is violated during inflation. 
 This can be realised if there are fields that acquire a {\it vev}
 depending on spatial coordinates, as scalars $\phi = \phi(x^i)$, or alternatively if there are fields that
 select a preferred direction, as vector configurations that break rotational invariance. 
Concrete realisations of both these possibilities can be found,
 for example,  in models where inflationary fields 
 acquire vacuum expectation values along space-like directions, motivated by  Solid Inflation \cite{Endlich:2012pz, Kang:2015uha,Gruzinov:2004ty} 
 or    inflationary set-ups involving vector fields (see e.g. \cite{vec, nonabel}). \\
 A general EFT approach allows us to study in a model
 independent way the consequences of this particular symmetry breaking. 
  The phenomenology of these models can be quite different with respect to standard
scenarios.  They can lead to a blue spectrum of gravity waves,  anisotropic features  in  non-Gaussianities,  and new couplings among different sectors (scalar-tensor-vector) of fluctuations 
(see e.g.~\cite{aniso-fAA}). Also, at the level of the background, scenarios that break   space diffeomorphisms
  can accommodate  models 
 that break  isotropy, possibly related to some of the anomalies in the CMB (see e.g. the recent review~\cite{Schwarz:2015cma}). Hence, it is interesting to develop a model independent approach for studying the breaking of space diffeomorphisms
during inflation, exploiting techniques based on the EFTI. This is the
scope of this work.  Preliminary studies have been made in \cite{Cannone:2014uqa, Cannone:2015rra} (see also \cite{Graef:2015ova,Lin:2015cqa} and the
recent \cite{Abolhasani:2015cve}). Here we further develop this subject, directly working with a Goldstone action for
fluctuations.
  When working at sufficiently high energies, we can expect that gravity decouples, and spatial diffeomorphisms reduce to global  space translations and rotations: the breaking
  of these symmetries lead again to Goldstone bosons. In particular, a scalar `phonon'  appears,
      that we call $\sigma$ and that
       is associated with the broken translational invariance. This
        Goldstone field $\sigma$ 
        interacts with the Goldstone boson $\pi$ associated 
  with  the breaking of time translations. Such couplings are constrained by non-linearly realised symmetries.
  They  lead to interesting effects, that
   we analyse for  the first time and
  that -- as we will explain -- 
   are not  obtained  in 
the  standard EFTI (i.e. where only time-reparameterisation invariance is broken) or Solid Inflation scenarios.

\smallskip

 Indeed, 
 we determine two broad  physical effects that are distinctive of our set-up, that we analyze in  detail:
 
 \begin{itemize}
 \item
  The first  is specific of the scalar sector, and exploits the new couplings between the two scalar Goldstone
 modes of broken symmetries. We find potentially  large 
 contributions to inflationary observables,
   that can give sizeable effects even in the limit of small breaking of space diffeomorphisms.
 Such contributions lead to a change in the  amplitude of the power spectrum of scalar fluctuations, and, more interestingly,
 direction dependent contributions to the squeezed limit of the scalar and tensor bispectra (in the sense that the bispectra depend
 non trivially on the angle between the three wavevectors, and can be parametrized with Legendre polynomials $P_{\rm L}$ and amplitude coefficients $c_{\rm L}$ as in~\cite{Shiraishi:2013vja}).

  We discuss the physical
  consequences of 
  these findings, pointing our similarities and differences with  previous results in the literature, as Solid Inflation \cite{Endlich:2012pz},  inflationary models involving vector \cite{Bartolo:2012sd, Bartolo:2013msa}  or higher spin  \cite{Arkani-Hamed:2015bza} field components.
  \item The second effect is instead more specific of tensor sector, and exploits novel possibilities
  for   tensors to couple with themselves and with  scalars. Such possibilities  are associated with  operators that
  are allowed only if we break also  space reparameterisation invariance during inflation.  They can lead to a blue
  spectrum for gravitational waves, and
 moreover to 
 a  particular structure for the squeezed limit of tensor-scalar-scalar bispectra, that violates single field consistency
 relations and can lead to distinctive observable signatures.
 \end{itemize}
 
 We discuss these findings in  Sections \ref{sec-sys}-\ref{sec-3ptf}, and we conclude in Section \ref{sec-concl}. In the three Appendixes we give technical details about the decoupling limit and the strong coupling scale in our model and we estimate the size of a galaxy survey in order to give informations on the pattern of symmetries of our model.

 \section{System under consideration}\label{sec-sys}

General Relativity is 
 built in terms of a powerful symmetry principle, diffeomorphism invariance:
  \be\label{difdef}
 x^\mu \to  x^\mu +\xi^{ \mu}( x^\nu)\,.
 \ee 
  Such symmetry can be spontaneously broken by sources 
 contributing to the energy momentum tensor that explicitly depend on the coordinates. 
  For example, an homogeneous scalar field $\phi(t)$ breaks time reparameterization invariance, 
   since its value is generally not preserved under the transformation
    $\phi(t+\xi^0(x^\mu))\neq\phi(t)$.
  In this paper, we study the case of a system
 that spontaneously breaks  {\it all} diffeomorphisms during inflation. 
  The spontaneous breaking of diffeomorphisms implies that perturbations
  of 
   symmetry-breaking fields
     transform non-linearly under diffeomorphisms.  
   Such situation can be achieved if fields select preferred  spatial directions, or depend explicitly on space-like coordinates.
  
  The study 
   of this system can  be 
   carried on following  different approaches, that we now briefly discuss. The first approach consists on working in what is called the `unitary gauge'. 
  One makes the hypothesis  that 
   the  system breaks diffeomorphism symmetries in such a way that 
  a gauge can be selected, where the fluctuations of the fields
 sourcing 
 inflation can be set to zero, and perturbations  are stored in the metric only \footnote{Let
  us emphasise that this condition is  not automatically satisfied in all models of inflation. Consider a
  system of two-fields inflation, $\phi_i$ with $i\,=\,1,\,2$, where both fields contribute to inflation
   acquiring a time-dependent {\it vev}. Their perturbations transform non-trivially under time reparameterization, $\delta \phi_i\to
   \delta \phi_i+\partial_t {\phi}_i\,\xi^0$. Having a unique function $\xi^0$ to play with, we don't have enough freedom
   for setting both $\delta \phi_i$ to zero.  
  }. This gauge choice makes
 the counting of the degrees of freedom particularly simple, and provides a geometrical interpretation
 of the dynamical fluctuations. 
     The possibility of making this gauge choice requires  that we 
 can work with 
  at most four  fields, that acquire vacuum expectation values spontaneously 
  breaking the symmetry. We  label them as  $\phi^\mu$, $\mu=0,..,3$. We then assume that
     their own   perturbations can be
 set to zero appropriately selecting the four functions $\xi^\mu$ in eq \eqref{difdef}. 
  This is our definition of unitary gauge (a similar condition was studied in \cite{Endlich:2012pz}).
  In this gauge, the dynamical degrees of freedom are stored in the metric: the usual transverse, plus
  all the {\it longitudinal} polarisations of the graviton. The resulting theory can be seen as an effective theory of (Lorentz
  violating) massive gravity in a cosmological space-time \cite{Rubakov:2004eb,Dubovsky:2004sg}.  
Besides the two transverse helicities, the longitudinal graviton polarisations  can account for at most four more degrees of freedom:
two form a transverse
vector and two are scalars. Notice that the scalars can both have healthy dynamics around
a cosmological space-time (i.e. one of them does not necessarily correspond to a ghost, as in flat space \cite{Blas:2009my}).

While the unitary gauge is 
 well suited for   geometrically understanding the dynamical   degrees of freedom, in our work 
  we  adopt a second  
  approach to study an inflationary  system with broken space-time diffeomorphisms. 
   We interpret the new dynamical
  modes that  arise  as Goldstone bosons of broken
 space-time symmetries.  
In order to do so,
  it is convenient to define our coordinates to be aligned with the background values of the fields that spontaneously break diffeomorphisms.

  The vacuum expectation values for the symmetry breaking fields are
 \begin{equation}\label{intro:vacuum}
  \bar\phi^0 = t\;, \qquad \bar\phi^i = \alpha\, x^i \;,
 \end{equation}
 $\bar \phi^0$ and $\bar \phi^i$ are respectively clock and rulers during inflation.  
 The parameter $\alpha$ controls the breaking of spatial diffeomorphisms: we assume it to be small,  and we will use it as an expansion parameter.
Using the St\"uckelberg trick, we can restore full diffeomorphism invariance by introducing a set of four fields, $\pi$ and $\sigma^i$, and write
the gauge invariant combinations $\phi^\mu$ as
\begin{equation}\label{intro:cpl}
  \phi^0 = t+\pi\;, \qquad 
  \phi^i = \alpha\, x^i+\alpha\,\sigma^i \;.
 \end{equation}
The St\"uckelberg fields $\pi$ and $\sigma^i$ transform under diffeomorphisms such to render the previous combinations
 gauge invariant. 
 For the system that we consider, $\sigma_i$ can be decomposed into longitudinal $\sigma_{L}$
 and transverse components $\sigma^{T}_i$. The
 longitudinal component $\sigma_{L}$  interacts with $\pi$, starting already at quadratic level:  the interaction among these scalars  will be the main topic of our work.

 We make further assumptions: we would like to preserve homogeneity and isotropy, imposing  extra internal symmetries  on the field configuration \cite{Endlich:2012pz},
 \begin{equation}\label{intro:internal}
  \phi^i\to O^i_j\phi^j\;,\qquad \phi^i\to\phi^i+c^i\;, 
 \end{equation}
 where $O^i_j \in SO(3)$.
 We further assume an approximate shift symmetry $\phi^0\to\phi^0+c^0$, which is a technically natural assumption to protect the small
 time dependence of the coefficients that will appear in the action.  
 Notice that  these internal symmetries we impose act on field space. Diffeomorphism invariance of eq \eqref{defdi} acts
  on coordinate space instead, and is spontaneously
 broken in our system. Notice that our pattern of symmetry breaking is different from the recent \cite{Abolhasani:2015cve}, that breaks
  rotational symmetry in the internal field
 space.

 With this in mind, we can write --  at lowest order in a 
 derivative expansion -- the diffeomorphisms  invariant action
  describing our system
 \begin{equation}\label{intro:action}
  S = \int \dif^4 x \sqrt{-g}\left[\frac{1}{2}\mpl^2 R + F(X,Y^i,Z^{ij})\right] \;,
 \end{equation}
 where $F$ is an arbitrary function,
 respecting the internal group of spacetime shifts and rotations \eqref{intro:internal} and $g$ is the determinant of the metric tensor. The building blocks
 that appear in the function $F$ 
  are the operators:
 \begin{eqnarray}\label{intro:blocks}
  X      & = & \de_\mu\phi^0\de_\nu\phi^0g^{\mu\nu} \;, \nonumber\\
  Y^i    & = & \de_\mu\phi^0\de_\nu\phi^ig^{\mu\nu} \;, \\
  Z^{ij} & = & \de_\mu\phi^i\de_\nu\phi^jg^{\mu\nu} \;, \nonumber
 \end{eqnarray}
 where $i=1,2,3$.
In what follows, we discuss the consequences of this form of the action for the dynamics of the St\"uckelberg fields.

 \section{Inflationary background and fluctuation dynamics}

    \subsection{The equations for the background}

     Our first task is to determine
      the background evolution. We selected the background values for the 
      fields that break diffeomorphisms to be aligned with the  space-time  coordinates, as in eq. \eqref{intro:vacuum}. 
       Such background  fields are expected to  drive inflation.
      We now
      consider    
      what conditions our function $F$ 
    have to satisfy,  in order to generate a quasi-de Sitter period of inflationary expansion. We start assuming 
     Friedmann-Robertson-Walker ansatz for the metric
     \be
     g_{\mu\nu}=\mbox{diag}(-1,a^2,a^2,a^2)\,,
     \ee
   where $a$ is the scale factor of the universe.  The energy-momentum tensor of our theory reads:
     
 \begin{equation}
  \begin{array}{lcl}
   T_{\mu\nu} & = & \displaystyle-\frac{2}{\sqrt{-g}}\frac{\delta S}{\delta g^{\mu\nu}} = \\
              & = & \displaystyle  g_{\mu\nu}F-2\left(F_X\,\de_\mu\phi^{0}\de_\nu\phi^{0}+F_{Y^i}\,\de_\mu\phi^{0} \de_\nu\phi^{i}
		    +F_{Z^{ij}}\,\de_\mu\phi^{i}\de_\nu\phi^{j}\right) \;,
  \end{array}
 \end{equation}
 where  the subscripts of $F$ stand for the partial derivatives with respect to the operators
 \eqref{intro:blocks}. When computed on the background
 values of the fields, eq \eqref{intro:vacuum}, Einstein equations lead to the
   Friedmann equations (where $H=\dot{a}/a$ and overlines denote quantities evaluated on the background):
 \begin{eqnarray}
  3\mpl^2H^2     &=& \left(-\bar{F}-2\bar{F}_X\right) \label{friedmann1}\;,\\
  -2\mpl^2\dot H &=& -2\left(\bar{F}_X+\frac{\alpha^2}{a^2}\bar{F}_Z\right)\;.\label{friedmann2}
 \end{eqnarray}
 On the background, the operators \eqref{intro:blocks} read 
 \begin{equation}\label{bvo1}
  \bar X = -1 \;,\qquad \bar Y^i=0\;,\qquad \bar Z^{ij}=\frac{\alpha^2\delta^{ij}}{a^2(t)} \;.
 \end{equation}
 Notice that $ Z^{ij}$ depends on $\alpha$ -- being associated with the breaking of space diffeomorphisms --
  but  it also 
  explicitly depends on time, through the scale factor.

 The isotropy of the background requires
 \begin{equation}
  \bar{F}_{Z^{ij}} = \bar{F}_Z \delta_{ij} \;,\qquad \bar{F}_{Y^i}=0\;.
 \end{equation}
 Our configuration solves all the background equations of motion (included the ones associated with the fields $X$, $Y^i$, $Z^{ij}$)
  if the following condition is satisfied \footnote{Notice that this equation is equivalent to the continuity equation. In the limit $\alpha\to 0$, one consistently obtain a limit $F_X \to 0$, which is the limit
of the continuity equation for a $F(X)$ theory when the symmetry $\phi_0\to\phi_0+c$ is taken as an exact symmetry.}:
 \begin{equation}
  2\alpha^2\bar F_{XZ} = a^2 F_X \;.
 \end{equation}
Using this information, we can express  the slow-roll epsilon parameter $\epsilon=-\dot{H}/H^2$  as
 \begin{equation} \label{epsilon}
  \epsilon = \frac{3\bar X \bar  F_X - \bar Z \bar F_Z}{-\bar F + 2\bar X \bar F_X} \;.
 \end{equation}
 To obtain a phase of  inflation   we require $\epsilon\ll1$, which, barring accidental cancellations,
 can be naturally obtained if the function $F$ has only a weak dependence on both $X$ and $Z$:
 \begin{equation}
\left(  \frac{\dif \log F}{\dif \log X}, \,\,\frac{\dif \log F}{\dif \log Z}\right)\,\ll\,1\,.
 \end{equation}
 Physically, the slow-roll parameter $\epsilon$ is  
 associated with the `ticks' of the inflationary clock. A small $\epsilon$
  is associated with a configuration characterised by 
     extremely slow ticks of the clock, corresponding to   
   a quasi-de Sitter space-time.
 The rhythm of the inflationary clock  ticks also varies, and is controlled by a
  second, independent slow-roll parameter $\eta\,=\,\dot{\epsilon}/(\epsilon\,H)$. A small  $\eta$ ensures
   that  
     changes in the rate of the inflationary clock 
    occur slowly, so 
   to provide 
  a sufficiently long period of inflation. The condition 
   $|\eta|\ll1$  requires
 \begin{equation}\label{eta}
  \left| \eta = 2\epsilon + \frac{6\bar F_{XZ}+2\bar Z\bar F_Z+2\bar Z^2\bar F_{Z^2}}{-3\bar F_X-\bar Z\bar F_Z}\right|\ll1
\,. 
 \end{equation}
Having a small value for $\epsilon$ implies,
 at leading order in slow-roll,  that  the quantities
\be
\frac{F_X}{F} {\text{   \hskip1cm  and  \hskip1cm}} \frac{F_Z}{a^2\,F}\;, 
\ee
are constant that do not  depend on space-time coordinates (notice the explicit presence of the scale factor in the second one).
   
 Our set-up does  not correspond to a single clock model of inflation. 
  We can identify
   two  independent contributions  that control the inflationary clock.  
 The first  is associated with the breaking of time reparameterisation, through  an explicitly time-dependent
  background value for the field $\bar \phi^0$,   as in eq \eqref{intro:vacuum}.  
 The second is related with the time dependence of the quantity $\bar Z^{ij}$ introduced in eq \eqref{bvo1}. 
  $\bar Z^{ij}$ is associated with the breaking
 of space diffeomorphisms, and is defined
  in terms of 
  the inflationary rulers $\bar \phi^i$ in eq \eqref{intro:vacuum}.
 $\bar Z^{ij}$  
  acquires a dependence on  the scale factor 
  $a(t)$ (due to the contraction with  the spatial part of the metric). A similar fact is found also in Solid Inflation \cite{Endlich:2012pz}.
   
   These two  contributions to the energy momentum tensor both 
   independently  control the inflationary clock. Hence we are not dealing
    with a purely adiabatic system.   
   And indeed, we will see next that
    we can identify two dynamical 
     scalar fluctuations around our background configuration, each corresponding to
    a  Goldstone boson of a different broken symmetry. The 
     non-adiabatic properties of our set-up are quite distinctive though, and are the topic of the remaining
     discussion.

\subsection{Quadratic action for St\"uckelberg  fields} 
 
We now discuss the structure of quadratic fluctuations of the transverse components of the metric,
  and the St\"uckelberg fields 
$\pi$, $\sigma^i$  introduced in eq \eqref{intro:cpl} as 
 \begin{equation} \label{goldstoneexpansion}
  \phi^0 = t + \pi \;,\qquad \phi^i = \alpha( x^i + \sigma^i)\;,
 \end{equation}
 as 
  fields restoring diffeomorphism invariance.

 In principle, besides the (self-)interactions of $\pi$ and $\sigma^i$, also 
  interactions  of these fields with the metric components $\delta g^{00}$, $\delta g^{0i}$, $\delta g^{ij}$
    should be taken into account. However, we can consider the theory at very short distances -- corresponding
    to   energy scales  $E=k/a\gg H$ --  where
     the
    effects of gravity backreaction can be neglected. 
    Gravitational modes decouple: the
      local diffeomorphisms of general relativity reduce to the global symmetries of Lorentz boosts
      and translations.   
     In this  decoupling limit the fields $\pi$ and $\sigma^i$ can be interpreted
    as Goldstone bosons of these broken global symmetries, and these degrees of freedom interact only with themselves \footnote{
    See Appendix \ref{app-A} for a technical discussion of decoupling limit in our set-up.}.

After these considerations, let us then focus on the system in a high energy decoupling limit, where the St\"uckelberg fields
can be  identified with Goldstone bosons of broken space-time diffeomorphisms. We 
 start writing 
 the quadratic actions for these systems  \footnote{The internal symmetries
 \eqref{intro:internal} limit the possible operators that can appear in the action. For example, deriving $F$ twice with respect
 to $Z^{ij}$ gives $\dif F/\dif Z^{ij}\dif Z^{kl} = F_{ZZ}\delta_{ik}\delta_{jl}+F_{Z^2}\delta_{ij}\delta_{kl}$.} 
at leading order in slow-roll parameters and the parameter $\alpha$,  neglecting 
   gravitational corrections in our decoupling limit:

 \begin{eqnarray}
  S^{(S)} &=&\int\dif^4x\,a^3\Bigg[\Bigg(-\bar{F}_X+2\bar{F}_{X^2}\Bigg)\dot\pi^2
	   +\Bigg(\bar{F}_X+\frac{\alpha^2\bar{F}_{Y^2}}{2a^2}\Bigg)\frac{\de_i\pi\de^i\pi}{a^2}+
	    \alpha^2  \Bigg(\frac{\bar{F}_{Y^2}}{2}-\bar{F}_Z\Bigg)\dot\sigma_L^2\nonumber\\
	  & & +\alpha^2\Bigg(\bar{F}_Z+\alpha^2\frac{2\bar{F}_{ZZ}}{a^2}+\alpha^2\frac{2\bar{F}_{Z^2}}{a^2}\Bigg)\frac{\de_i\sigma_L\de^i\sigma_L}{a^2}
	   +\alpha^2\frac{4\bar{F}_{XZ}}{a^2}\sqrt{-\nabla^2}\dot\pi\sigma_L-\alpha^2\frac{\bar{F}_{Y^2}}{a^2}\sqrt{-\nabla^2}\pi\dot\sigma_L\Bigg] \;, \nonumber\\\label{scalaraction}\\
  S^{(V)} &=& \int\dif^4x\,a^3\Bigg[ \left(\frac{\bar{F}_{Y^2}}{2}-\bar{F}_Z\right)\dot\sigma_T^i\dot\sigma_{T,i}
	  +\left(\bar{F}_Z+2\frac{\bar{F}_{ZZ}}{a^2}\right)\frac{\de_j\sigma_T^i\de^j\sigma_{T,i}}{a^2}\Bigg]\;,\label{vectoraction}\\
	  & & \nonumber \\
  S^{(T)} &=& \int\dif^4xa^3\frac{1}{8}\Bigg[\mpl^2\left(\dot\gamma_{ij}\dot\gamma^{ij}-\frac{\de_k\gamma_{ij}\de^k\gamma^{ij}}{a^2}\right)
	  +\alpha^2\left(\frac{\bar{F}_Z}{a^2}+\frac{\alpha^2\bar{F}_{ZZ}}{2a^4}\right)\gamma_{ij}\gamma^{ij}\Bigg]\;.\label{tensoraction}
 \end{eqnarray}
where $S,V, T$ represent the scalar, vector and tensor sectors respectively. The fields $\pi$ has dimension of inverse of mass,  and $\sigma$ is dimensionless. 
 The field $\sigma^{i}$ has been decomposed in
  a (vector)  transverse component and 
  a (scalar) longitudinal one:
 \begin{equation}
  \sigma^i=\sigma_T^i+\frac{\de^i\sigma_L}{\sqrt{-\nabla^2}}\;.
 \end{equation}
 As explained above, when 
  all diffeomorphisms are broken, in general six degrees of freedom are dynamical: two scalar fluctuations, the two components of a
 transverse vector, and the two helicities of a traceless transverse tensor.

The most evident
 consequence of our set-up is that we now have a system of two interacting scalars, $\pi$ and $\sigma$,
 Goldstone bosons of two different symmetries.  These scalars  are coupled through distinctive derivative 
 operators, controlled by the pattern of symmetry breaking in our system. Notice that masses, and
 non-derivative couplings among the fields, do not arise at our level of approximation, because of the symmetries~(\ref{intro:internal}) and since we are neglecting gravitational effects.   Our action is different with respect to the EFTI action for multifield inflation \cite{Senatore:2010wk}.
 While the previous actions are quadratic in perturbations, also non-linear interactions
 can be included: we will do so
 in Section \ref{sec-3ptf}.

 We should also  check that the actions \eqref{scalaraction}, \eqref{vectoraction} and \eqref{tensoraction} do not lead
 to dangerous instabilities. For example, the coefficient of the time kinetic operators should have the right sign:
 \begin{eqnarray}
  -\bar F_X+2\bar F_X^2  & > & 0\;, \\
  \bar F_{Y^2}-2\bar F_Z & > & 0\;.
 \end{eqnarray}
 At the same time one should impose that the speeds of sound,
 \begin{eqnarray}
  c_\pi^2    &=& \frac{\bar F_X + \bar \alpha^2 F_{Y^2}/2a^2}{\bar F_X -2\bar F_{X^2}}\;, \label{cpi}\\
  c_\sigma^2 &=& \frac{\bar F_Z + 2\alpha^2\bar F_{ZZ}/a^2+2\alpha^2\bar F_{Z^2}/a^2}{F_Z-F_{Y^2}/2}\;, \label{csigma}\\
  c_T^2      &=& \frac{\bar F_Z+2\alpha^2\bar F_{ZZ}/a^2}{\bar F_Z + \bar F_{Y^2}/2}\;,
 \end{eqnarray}
 lie in the interval $0<c_s^2\le 1$, where $s\equiv \pi, \sigma, T$.
 The complete list of relations between the coefficients that one can derive is not particularly
 illuminating, but we  checked that  there are regions of the parameter space where there are no dangerous instabilities.
 The allowed range of  parameters will of course be important when trying to compare with cosmological observations,
 but this topic goes beyond the scope of the present work. 
 Moreover, to avoid excessive time evolution for these quantities
  during inflation, 
  we can impose that the ``slow-roll''
 parameter associated with the speeds of sound should be small:
 \begin{equation} \label{s1}
  s_c = \frac{\dot c_s}{c_s H}\ll1\;.
 \end{equation}
 This  again can give constraints on combinations of parameters in the action, when comparing with observations.
 { Moreover it suggests that, like $\bar F_X$ and $\bar F_Z/a^2$, also the other coefficients in the action, like for example
 $\bar F_{Y^2}/a^2$ are slowly varying and can be taken as constant. It would be interesting to see what are the consequences
 of relaxing this assumption and consider non-trivial time dependencies.}

  \subsection{The expression for the curvature perturbation} \label{curvpert-sec}

          There are two commonly used gauge invariant definitions of curvature perturbations, the curvature
     perturbation on uniform density hypersurfaces, $\zeta$, and the comoving curvature perturbation $\cal R$.
     In single field inflation, on superhorizon scales
 these quantities are conserved, and coincide up to a sign: 
  see e.g. \cite{Bassett:2005xm} for
 a review. 
      
      This implies that     
      any 
      result obtained in the aforementioned  sub-horizon, decoupling limit remains valid also at  superhorizon
      scales, since  the curvature perturbation gets frozen there in single-field inflation. As we explained
      above, our system is strictly speaking not single field, and non-adiabatic contributions can arise. They are controlled
      by a small quantity though -- the parameter $\alpha$ that characterises the breaking of space diffeomorphisms, see
       eq \eqref{intro:vacuum}. 
        The expression for the comoving curvature perturbation in the decoupling limit reads for our
        system\footnote{In the flat gauge the comoving curvature perturbation is defined as $\mathcal{R}=H\delta u$,
        where $\delta u$ is the longitudinal component of the perturbed $4$-velocity of the fluid \cite{Malik:2008im}.}:
        \begin{equation}\label{curvature}
  \mathcal{R} = \frac{H}{(-\mpl^2\dot H)}\left[\left(-\bar F_X+\frac{\alpha^2\bar F_{Y^2}}{2a^2}\right)\pi
  + \alpha^2\left(2\bar F_Z-\bar F_{Y^2}\right)\frac{\dot\sigma_L}{\sqrt{-\nabla^2}}\right]\;.
 \end{equation}

       In the limit of $\alpha$ small, this expression reduces to the `single-field' expression 
       \be{\cal R}\,=\,-H\,\pi\;,
       \label{reft}
       \ee
        commonly 
       used in EFTI \cite{Cheung:2007st} (where ${\cal R}$ is dubbed $\zeta$). In what follows, we will work in a small $\alpha$ limit, so 
       that the definition \eqref{reft} is sufficiently accurate, and we can neglect its time-dependence at
       superhorizon scales. This is also justified because, nevertheless, we will
        find interesting potentially sizeable corrections to the $n$-point functions for ${\cal R}$ associated with
           the complete breaking of diffeomorphism invariance.

  \section{The two-point functions }

 \subsection{The power spectrum for scalar fluctuations}\label{sec-pssf}

 In this section we consider the consequences of the new symmetry pattern in the second order action
 of the scalar perturbations.
 First, let us rewrite the action \eqref{scalaraction} in terms of the normalized fields $\hat{\pi}$ and $\hat{\sigma}$,
 \begin{equation} \label{canonical_pi}
  \hat{\pi} = \sqrt{2\left(-\bar F_X +2\bar F_{X^2}\right)}\pi \;,\qquad {\hat{\sigma}} = \alpha\sqrt{2\left(\frac{\bar F_{Y^2}}{2a^2}-\frac{\bar F_Z}{a^2}\right)} \sigma_L\;,
 \end{equation}
 \begin{equation}\label{canonicalaction}
  S^{(S)} = \int\dif^4x\,a^3\Bigg[\frac{1}{2}\left(\dot{\hat{\pi}}^2-c_\pi^2\frac{\de_i{\hat{\pi}}\de^i{\hat{\pi}}}{a^2}\right)+
	      \frac{1}{2}a^2\left(\dot{\hat{\sigma}}^2-c_\sigma^2\frac{\de_i{\hat{\sigma}}\de^i{\hat{\sigma}}}{a^2}\right)
	      +\alpha\lambda_1\sqrt{-\nabla^2}\dot{\hat{\pi}}{\hat{\sigma}}+\alpha\lambda_2\sqrt{-\nabla^2}{\hat{\pi}}\dot{\hat{\sigma}}\Bigg] \;,
 \end{equation}
 where the speeds of sound are written in eqs. \eqref{cpi}, \eqref{csigma} and
 \begin{eqnarray}
  \lambda_1 &=& \frac{2\bar F_{XZ}/a^2}{\sqrt{\left(-\bar F_X +2\bar F_{X^2}\right)\left(\bar F_{Y^2}/2a^2-\bar F_Z/a^2\right)}} \label{lambda1}\:, \\
  \lambda_2 &=& \frac{-\bar F_{Y^2}/a^2}{2\sqrt{\left(-\bar F_X +2\bar F_{X^2}\right)\left(\bar F_{Y^2}/2a^2-\bar F_Z/a^2\right)}} \label{lambda2}\;.
 \end{eqnarray}
The normalization of the fields have been defined so that the parameters $\lambda_1$, $\lambda_2$
are constant at leading order in slow-roll. The price to pay is that we leave an explicit 
factor of $a^2$ in front of the `kinetic action' for the Goldstone mode $\sigma$ in eq \eqref{canonicalaction}.

 We can see that, for small values of the parameter $\alpha$, the interaction terms that mix the two fields can be treated
 as perturbations on top of a free Lagrangian for the two scalars involved. Thanks to this fact,
 we can 
  perturbatively 
  compute the spectrum for the fluctuation $\pi$ by the following  procedure. First, we evaluate it at zero order in the  parameter $\alpha$.  
  Then, we 
 compute perturbative corrections in $\alpha$, using the in-in formalism. This calculation will provide a quantitative way to evaluate
 how  the second Goldstone boson $\sigma$ affects the properties of the two-point function of $\pi$ and the curvature
 perturbation. 
  Physically, we are interested to this question because we have learned that the 
 contribution to   curvature perturbation ${\cal R}$ is 
    mostly due
      the  field $\pi$, in the limit of small values for $\alpha$ (see section \eqref{curvpert-sec}). On the other hand  
      we will learn that contributions of $\sigma$ to two and higher point functions of ${\cal R}$ can
       be sizeable even in the limit of small $\alpha$.

Let us then proceed computing the power spectrum for $\pi$.  
 The zeroth order power spectrum is  straightforward to obtain:
 \begin{equation}
  \langle\hat{\pi}_{\vec{k}_1}\hat{\pi}_{\vec{k}_2}\rangle =  (2\pi)^3\delta(\vec{k}_1+\vec{k}_2)\,\frac{2\pi^2}{k_1^3}\hat{\mathcal{P}}_0 \;,
 \end{equation}
 where
 \begin{equation}
  \hat{\mathcal{P}}_0 = \frac{H^2}{4\pi^2c_\pi^3} \;.
 \end{equation}
 Using the normalization coefficient \eqref{canonical_pi}, the power spectrum of the original fields $\pi$ reads:
 \begin{equation}
  \hat{\mathcal{P}}_0 = \frac{H^2}{8\pi^2c_\pi^3(-\bar F_X + 2\bar F_{X^2})}
  = \frac{H^2}{8\pi^2\, c_\pi \left(-\bar F_X-\alpha^2\bar F_{Y^2}/2a^2\right)}\;,
 \end{equation}
 where we used the definition of the speed of sound \eqref{cpi}. Taking $\alpha\ll1$ and using eq. \eqref{friedmann2}, this result
 reduces to the standard result of single-field inflation with only time-diffeomorphism breaking, as in that case
 $\bar F_X = \mpl^2\dot H$:
 \begin{equation}
\hat{\mathcal{P}}_0  \left(\alpha\ll 1 \right)\, 
 =\, \frac{H^4}{8\pi^2\, c_\pi \left(-\bar F_X\right)} = \frac{H^2}{8\pi^2 \mpl^2\epsilon \,c_\pi}\;.
 \end{equation}

\begin{figure}[t!]
 \begin{center}
  \includegraphics[scale=0.22]{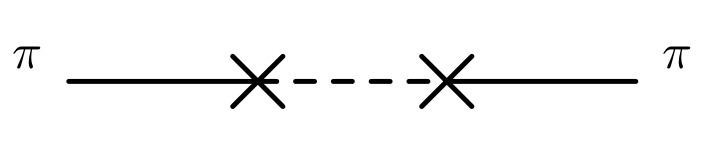}
\caption{Leading diagram for computing the symmetry breaking contributions to   $\left\langle \pi^{2}\right\rangle$.}
\label{massinsertion1}
 \end{center}
\end{figure}

  The effect of the interaction terms in the second order
 action \eqref{canonicalaction} can be now computed   using the in-in formalism \cite{Maldacena:2002vr, Weinberg:2005vy}.
 The leading correction to the power spectrum is given by
 \begin{eqnarray}
  \delta \langle \hat{\pi}_{{\vec k}_1} \hat{\pi}_{{\vec k}_2  } \left( \tau \right)  \rangle 
  & = & - \,  \int_{\tau_{{\rm min}}}^\tau d \tau_1  \int_{\tau_{{\rm min}}}^{\tau_1} d \tau_2 
  \left \langle \left[ \left[ \hat{\pi}_{{\vec k}_1}^{(0)}  \hat{\pi}_{{\vec k}_2  }^{(0)} \left( \tau \right)  ,  {\cal H}_{{\rm int}}^{(2)} \left( \tau_1 \right) \right] ,\,     {\cal H}_{{\rm int}}^{(2)} \left( \tau_2 \right) \right] \right \rangle \;,
  \label{inin-zz}
 \end{eqnarray}
 where ${\cal H}_{\rm int}^{(2)}$ is the second order interaction hamiltonian and $\tau_{\rm{min}}$
 corresponds to the time at which the contribution of the long mode starts to become dominant. This correction 
  is represented as mass insertion diagram in Fig \ref{massinsertion1}. 
  In our case, from \eqref{canonicalaction}, we have:
 \begin{equation}\label{secondint}
  \mathcal{H}_{{\rm int}}^{(2)} \left( \tau \right)=\mathcal{H}_{{\rm int},1}^{(2)} \left( \tau \right)+\mathcal{H}_{{\rm int},2}^{(2)} \left( \tau \right)\,,
 \end{equation}
 where
 \begin{equation}
  \mathcal{H}_{{\rm int},1}^{(2)} \left( \tau \right) = \frac{\alpha\lambda_{1}}{(H \tau)^3} \int \frac{d^3 k}{(2\pi)^{3}} \, |k| \hat\sigma_{\vec{k}}^{\left(0\right)}(\tau)\,{\hat\pi_{-\vec{k}}}'^{\left(0\right)}(\tau)\;,
  \label{Hint1}
 \end{equation}
 \begin{equation}
  \mathcal{H}_{{\rm int},2}^{(2)} \left( \tau \right) = \frac{\alpha\lambda_{2}}{(H \tau)^3} \int \frac{d^3 k}{(2\pi)^{3}} \, |k| \hat{\pi}_{\vec{k}}^{\left(0\right)}(\tau)\,\hat\sigma_{-\vec{k}}'^{\left(0\right)}(\tau)\,.
  \label{Hint2}
 \end{equation}
The field operators  can be expanded in terms of their Fourier modes 
 \begin{eqnarray}
  \hat{\pi}_{\vec{k}}&=&u_{k}\,a_{\vec{k}} +u_{k}^*\, a^\dagger _{-\vec{k}}\,,\nonumber\\
  \hat\sigma_{\vec{k}}&=&v_{k}\,b_{\vec{k}} +v_{k}^*\, b^\dagger _{-\vec{k}}\,,
  \label{mode}
 \end{eqnarray}
 where the creation and annihilation operators respect the commutation rules:
 \begin{equation}
  \left[a_{\vec{k}} ,a^\dagger _{-\vec{k}'}\right]=\left(2\pi\right)^{3}\delta^{(3)}(\vec{k}+\vec{k}')\,,\;\;\;\;\;\left[b_{\vec{k}} ,b^\dagger _{-\vec{k}'}\right]=\left(2\pi\right)^{3}\delta^{(3)}(\vec{k}+\vec{k}')\,,\,\,\,\,\left[a_{\vec{k}} ,b^\dagger _{-\vec{k}'}\right]=0\,.
 \end{equation}
 This is because the  eigenfunctions for the two scalar modes are the solution of the classical equations of motion, derived from the
 (free) action that can be read from \eqref{scalaraction}. For the field $\pi$ we have:
 \begin{equation}
  u_k''-\frac{2}{\tau}u_k'+c_\pi^2k^2 u_k = 0 \;,
 \end{equation}
 where we have used $aH=-1/\tau+\mathcal{O}(\epsilon)$. This has the standard solution (after choosing the Bunch--Davies
 vacuum and fixing the integration constants):
 \begin{equation}\label{pi_wave}
   u_{\vec{k}}(\tau) =  -\frac{H}{\sqrt{2c_{\pi}^3k^{3}}}(1+i k c_{\pi}\tau)e^{-i k c_{\pi}\tau}\,.
 \end{equation}
 In the same way one can write the equation of motion for the field $\sigma$:
 \begin{equation}
  v_k'' - \frac{4}{\tau}v_k' +c_\sigma^2k^2v_k = 0\;,
 \end{equation}
 whose solution is:
 \begin{equation}\label{sigma_wave}
  v_{k}(\tau) = -\frac{H^2}{\sqrt{2c_{\sigma}^5k^{5}}}(-3-3i k c_{\sigma}\tau+c_\sigma^2k^2\tau^2)e^{-i k c_{\sigma}\tau}\,.
 \end{equation}
 Notice that the vacuum wave configuration for the field $\sigma$,  \eqref{sigma_wave},
   is different with respect to 
  the vacuum configuration
  for  $\pi$, eq \eqref{pi_wave}. The difference is due to the presence of the scale factor $a^2$ in front of the 
  kinetic term for $\sigma$, in the quadratic action for fluctuations.

 Performing the commutators
 and plugging the interaction hamiltonian \eqref{Hint1} into in \eqref{inin-zz},
 we arrive to integrals like:
 \begin{eqnarray}
  \delta \langle \hat{\pi}_{{\vec k}_1} \hat{\pi}_{{\vec k}_2  } \left( \tau \right)  \rangle 
  & = &-  \frac{4 \alpha^2\lambda_{1}^2    }{H^6 } \, \rm{Re}\left[ \int_{\tau_{\rm min}}^\tau \frac{d \tau_1}{\tau_1^3}  \int_{\tau_{\rm min}}^{\tau_1} \frac{d \tau_2}{\tau_2^3}\,\,\, k^2\right.\times\nonumber\\
  & &\left.\left(v_{k_1}(\tau_{2})v_{k_1}^{*}(\tau_{1}) u'_{k_1}(\tau_{2})u_{k_1}^{*}(\tau)\left(u'_{k_1} (\tau_{1}) u_{k_1}^{*}(\tau)-c.c\right)\right)\right] \;.
 \label{first}
 \end{eqnarray}
 Together with this, there are also the integrals coming from the substitution of $\mathcal{H}_{{\rm int},2}$ \eqref{Hint2},
 proportional to $\lambda_2^2$ and the mixed contributions proportional to $\lambda_1\lambda_2$.
 These integrals are dominated by the contributions at the times when the modes are outside the horizon, as on sub-horizon scale
 the oscillatory phases in the mode functions suppress the result\footnote{The spurious   divergences in the UV disappear when slightly deforming the
 countour of the time integration in the imaginary direction: see  \cite{Maldacena:2002vr}.}.
 Setting $c_\sigma \simeq c_\pi$ for simplicity, we can take $\tau_{{\rm min}} = -1/c_\pi k$ and perform the integral analitically. 
 We find the main contribution in the large scale limit:
 \begin{equation}
  \frac{\delta\mathcal{P}}{\mathcal{P}_0}=\frac{3\alpha^2 \lambda_1 N_k (3 \lambda_2 -\lambda_1(3 N_k + 6 \gamma_E +11-\log(64))}{c_\pi^2}\;,
 \end{equation}
 where $\gamma_E$ is the Euler gamma and $N_k$ is the number of efolds from the time when the mode $k$
 exits the horizon until the end of inflation. Notice that in eq \eqref{first} the integrand contains a factor of $k^2$, associated with the derivative interactions
 among the modes $\pi$ and $\sigma$. This factor compensates for the non-standard form of the vacuum solution \eqref{sigma_wave} for $\sigma$, proportional to $k^{-5/2}$
 at small scales (and not to   $k^{-3/2}$ as it usually  happens), leading to a scale invariant correction to the power spectrum.

  One can see that even when the breaking of spatial diffeomorphisms is small,
 $\alpha\ll1$, the effect of the interaction between the field $\sigma$ and the field $\pi$ can still be sizeable, as it is enhanced
 by the e-fold number $N_k$. In the limit of large $N_k$, the dominant correction to the power spectrum scales as 
 \begin{equation}
  \frac{\delta\mathcal{P}}{\mathcal{P}_0}=-\frac{9\,\alpha^2\, \lambda_1^2\, N_k^2}{c_\pi^2}\;.
 \end{equation}
 This quantity can be non-negligible  even if $\alpha$ is small, since the product 
 $\alpha \, N_k$ can be sizeable (say of order one) being it enhanced by $N_k$. 
The logarithmic enhancement of the power spectrum has a similar
 behavior already met in other set-ups, see e.g. \cite{Bartolo:2014xfa} 
 or the review 
 \cite{Seery:2010kh}
   \footnote{Notice that, since we are {\it not} in single field inflation,
 these enhancement effects {\it cannot} be `gauged away' by a rescaling of the curvature perturbation. See e.g. \cite{Gerstenlauer:2011ti} or the reviews \cite{Tanaka:2013caa}. }.  
These are novel effects that we first point out in this paper, and are essentially due to the interplay between our two Goldstone bosons during inflation. They 
 are physical, and arise even if the breaking of space diffeomorphisms is small, since they are enhanced by the e-fold number.
  
 Since for small $\alpha$ the curvature perturbation ${\cal R}$ is proportional to $\pi$ through the simple relation \eqref{reft}, we can write the following modified expression
  to the power spectrum for ${\cal R}$, induced by a log-enhancement due to  the Goldstone boson $\sigma$:
  
  \be
  \mathcal{P}_{\mathcal{R}} 
\,\simeq  \,\frac{H^2}{8\pi^2 \mpl^2\epsilon \,c_\pi}\left(1- \frac{9\,\alpha^2\, \lambda_1^2\, N_k^2}{c_\pi^2}\right)\;.
  \ee
   
 Although for the case of scalar power spectrum the correction induced by
 our pattern for  breaking spatial diffeomorphisms  amounts to a change in the amplitude, for higher point functions
  we can have more relevant, direction dependent effects, as we will discuss in Section \ref{sec-3ptf}.

 \subsection{The power spectrum for tensor fluctuations}
 
 The effect of breaking spatial diffeomorphisms can have  interesting effects also in the tensor sector. This fact has been already explored in
  \cite{Cannone:2014uqa,Cannone:2015rra}. At the level of two-point functions involving tensor modes, the  main difference with the 
  standard case is associated with  the 
  possibility of assigning a non-vanishing mass to the tensors, since a mass  operator is allowed by the absence of diffeomorphism invariance.
The resulting set-up can then be considered as an effective theory of (Lorentz-violating) massive gravity during inflation. It would be interesting
to find a consistent, UV complete theory of massive gravity that allows us to have  a large graviton mass during cosmological inflation,  and a small
graviton mass after inflation ends.
 Nevertheless, in our approach  based on EFT we do not need to rely on the existence of any specific UV realisation, and simply work with the most
 general set of operators, order by order in a field expansion. 

 Normalizing the  tensor field as 
 $\gamma_{ij} = \sqrt{2}\hat{\gamma}_{ij}/\mpl$ the quadratic Lagrangian for the two polarization modes of the gravitational fields
 has the form:
 \begin{equation}
  \mathcal{L} = \frac{1}{4}\sqrt{-g}\left[\de_\mu \hat{\gamma}_{ij}\de^\mu \hat{\gamma}^{ij}
  - m^2\hat{\gamma}_{ij}\hat{\gamma}^{ij}\right] \;,
 \end{equation}
 where $m^2=\alpha^2(\bar F_{Z}+\alpha^2\bar F_{ZZ}/2a^2)/\mpl^2a^2$.

 To compute the power spectrum, one decomposes $h_{ij}$ into helicity modes,
 \begin{equation}
  \hat{\gamma}_{ij}=\int\frac{\dif^3k}{(2\pi)^3}\sum_{s}{\bf \epsilon}_{ij}^{(s)}\hat{\gamma}_{\vec{k}}^{(s)}\me^{\mi {\bf k}\cdot{\bf x}}\;,
 \end{equation}
 where $s=\pm$ is the helicity index and $\epsilon_{ij}$ is the polarization tensor. Making the redefinition $\hat{\gamma}_{\vec{k}}=h_{\vec{k}}/a$, the equation of motion (in
 conformal time $d \eta\,=\,d t/a(t)$ and neglecting slow-roll corrections) reads:
 \begin{equation}
  h_{\vec{k}}''+\left[k^2-\frac{1}{\tau^2}\left(\nu^2-\frac{1}{4}\right)\right]h_{\vec{k}} = 0 \;,
 \end{equation}
 where $\nu=9/4-m^2/H^2$. The generic solution (for real $\nu$) is \footnote{For imaginary $\nu$ one can define a new
 $\tilde \nu=\mi\,\nu$ and solve the differential equations in the same way. However in this case the power spectrum would be
 suppressed by the ration $H/m$ and fall rapidly on very large scales \cite{Chen:2009zp}.}:
 \begin{equation}\label{h_wave_mass}
  h_{\vec{k}} = \frac{\sqrt{\pi}}{2}\me^{\mi(\nu+1/2)\frac{\pi}{2}}\sqrt{\eta}\,H_\nu^{(1)}(k\eta) \;,
 \end{equation}
 where $H^{(1)}_\nu$ is the Hankel's function of the first kind. At this point one can easily find the tensor power spectrum
 \begin{equation}
  \mathcal{P}_T \simeq \frac{2H^2}{\pi^2\mpl^2}\left(\frac{k}{k*}\right)^{3-2\nu} \;,
 \end{equation}
 where:
 \begin{equation}
  n_T \simeq m^2/H^2 \;.
 \end{equation}
 Together with the standard $-2\epsilon$ contribution to $n_T$, which can be easily found taking into account the slow-roll
 dependence in the equation of motion, this shows a non-trivial behaviour of the tensor tilt \cite{Cannone:2014uqa}:
 \begin{equation}
  n_T = -2\epsilon + \frac{2}{3}\frac{m^2}{H^2}\;.
 \end{equation}

 As the $m^2/H^2$ contribution adds with a positive sign,
 if the mass of the tensor is large enough, then the spectrum could become blue \cite{Cannone:2014uqa, Graef:2015ova}.
 Moreover this would happen
 without  inconsistencies, since it would not violate the Null Energy Condition, which is related to a change of sign of $\dot{H}$
 \cite{Creminelli:2006xe}. The interplay between the negative contribution of $\epsilon$ given by the time-diffeomophism
 breaking part and the positive contribution given by the breaking of space diffeomorphisms is a non-standard feature of this
 particular symmetry pattern.
  Notice also that, even though massive tensors are not constant after horizon exit, their evolution
 is very small as it is controlled by the small parameter $\alpha$. Indeed, if we take the limit $\alpha\ll1$, tensor mass becomes
 completely negligible and we come back to the standard form of the tensor wave function:
 \begin{equation}\label{gamma_wave}
  \gamma_{\vec{k}} = \frac{H}{\mpl k^{3/2}}(1+\mi k \tau)\me^{-\mi k \tau} \;.
 \end{equation}
 So we find that breaking spatial diffeomorphisms provides qualitatively new  effects in the power spectrum of tensor fluctuations. Other
 interesting effects arise when studying the bispectrum, as we are going to see in the next section.

   \section{The three-point functions}\label{sec-3ptf}

In this section, we
examine non-linear, 
 cubic interactions among cosmological fluctuations. 
 In particular, we will study bispectra involving scalar and tensor fluctuations.
 The study of bispectra is conceptually important since their squeezed limits are very informative for what concern general features of the physics driving inflation. 
For example, it is known that in models with adiabatic fluctuations only,  appropriate squeezed limits of three-point functions involving scalars and tensors, e.g. $\langle {\cal R}^3\rangle$ and  $\langle \gamma\,{\cal R}^2\rangle$,
 are related to the tilt of the scalar power spectrum \cite{Maldacena:2002vr, Creminelli:2004yq}.  
 
 When  non-adiabatic interactions are turned on, these consistency relations are violated in a way that depends on the model one considers.
 We are interested in understanding general features
  of how  
 the breaking of space-time diffeomorphisms affects the squeezed limits of three-point functions. We find  that the breaking of such symmetry leads to (tunable) quadrupolar contributions to these quantities (corresponding to $c_{\rm L=2}$ contributions in the parameterisation of~\cite{Shiraishi:2013vja}) besides ``pure'' local (monopole $c_{\rm L=0}$) contributions in the squeezed limit.
 
 Similar results have been already found in specific models, as Solid Inflation or models with vector fields \cite{Endlich:2012pz, Kang:2015uha,Shiraishi:2013vja,Bartolo:2012sd}, but our  EFT approach allows to generalize these results 
 and understand them as due to a specific pattern of symmetry breaking. 
 
 As done in the case of the power spectrum, we are mostly interested on operators that are specifically associated with the simultaneous breaking of time {\it and} space reparameterization
 invariance, since these operators can lead to effects that have not been studied so far, when breaking separately time \cite{Cheung:2007st} and space \cite{Endlich:2012pz} diffeomorphisms. Moreover, such effects can be sizeable, rendering them physically interesting even in a limit of small $\alpha$, the parameter associated with the
   breaking of spatial diffeomorphisms.

Given these motivations, the operators that we consider are specific of our construction that simultaneously break
space and time diffeomorphisms.  Up to second order in the parameter $\alpha$ they are the following:

 \begin{equation} \label{bisFXZ}
  \frac{8 \alpha^2\bar F_{X^2Z}}{3a^2} \dot\pi^2\de_i\sigma^i\;,
 \end{equation}
 \begin{equation} \label{bisFY}
  \alpha^2\bar F_{Y^2}\left(\dot\pi\dot\sigma^i\dot\sigma_i -\frac{\dot\pi\dot\sigma^i\de_i\pi}{a^2}+\frac{\gamma_{ij}\dot\sigma^i\de^j\pi}{a^2}
  -\frac{\gamma_{ij}\de^i\pi\de^j\pi}{a^4}-\frac{\dot\sigma^i\de_j\sigma_i\de^j\pi}{a^2}+\frac{\de_j\sigma_i\de^i\pi\de^j\pi}{a^4}\right)\;,
 \end{equation}
 \begin{equation} \label{bisFYX}
  \frac{2}{3}\alpha^2\bar F_{Y^2X}\left(-\dot\pi\dot\sigma^i\dot\sigma_i+\frac{2\dot\pi\dot\sigma^i\de_i\pi}{a^2}\right)\;.
 \end{equation}
  Here we are mostly interested in exploring interesting phenomenological consequences
 of our approach. On the other hand, the analysis of interactions can also be theoretically important to estimate the {\it strong coupling scale} at which unitarity
 bounds are violated in scattering experiments. We do not discuss  this argument in the main text, but we 
develop  it in Appendix \ref{app-B}.

 \subsection{The bispectrum for scalar fluctuations}

We start discussing  how 
 the operators breaking simultaneously space and time 
 diffeomorphisms affect the squeezed limit of the curvature three-point function.
The bispectrum of the curvature perturbation is defined as:
 \begin{equation}
  \langle\mathcal{R}(\vec{k}_1)\, 
  \mathcal{R}(\vec{k}_2)\, 
  \mathcal{R}(\vec{k}_3)
   \rangle = (2\pi)^3\delta^3(\vec{k}_1+\vec{k}_2+\vec{k}_3)\mathcal{B}(\vec{k}_1,\vec{k}_2,\vec{k}_3)\;.
 \end{equation}
  As for the two-point function, we  compute the contributions of the symmetry breaking operators 
    using a perturbative approach based on the in-in formalism. We take the quantity $\alpha$ controlling the  breaking 
    of space diffeomorphisms as a perturbation  parameter. In the limit of small $\alpha$,
    the curvature perturbation is proportional to the Goldstone mode $\pi$, up to small corrections, and
    is conserved on superhorizon scales: see the 
    discussion in Section \ref{curvpert-sec}, and the expression \eqref{reft} for the curvature perturbations.

 \smallskip
 
\begin{figure}[t!]
 \begin{center}
  \includegraphics[scale=0.22]{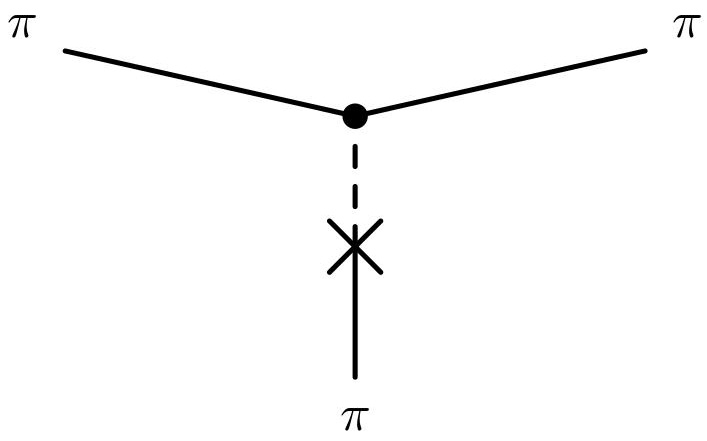}
  \caption{Leading diagram for computing the  symmetry breaking contributions to    $\langle\mathcal{R}^3\rangle$.}
\label{fig:bispectrum} 
 \end{center}
\end{figure}

 We then start with computing the  contribution to the tree level bispectrum of the Goldstone $\pi$, due to the mixing with the Goldstone $\sigma$.
We consider the contribution associated with the diagram represented in Fig. \ref{fig:bispectrum}. In the limit of
 small $\alpha$, the operators that we  consider are associated with a mass insertion  second order
  hamiltonian  $\mathcal{H}^{(2)}$, 
 given by \eqref{secondint}, and by the third order operator
 \begin{equation}\label{operators_bispectrum}
  \frac{\alpha}{\sqrt{2\left(\bar F_X+2\bar F_{X^2}\right)}}\Bigg[
  (\lambda_2+\lambda_3)\frac{\dot{\hat{\pi}}\de^i\dot{\hat{\sigma}}\de_i{\hat{\pi}}}{\sqrt{-\nabla^2}}
  -\lambda_2\frac{\de_j\de_i{\hat{\sigma}}\de^i{\hat{\pi}}\de^j{\hat{\pi}}}{a^2\sqrt{-\nabla^2}}
  -\lambda_4 \dot{\hat{\pi}}^2\sqrt{-\nabla^2}{\hat{\sigma}}\Bigg]\;,
 \end{equation}
that we express in terms of normalized fields \eqref{canonical_pi}.
The new parameters $\lambda_3$ and $\lambda_4$ are defined as
 \begin{eqnarray} 
  \lambda_3 &=& \frac{4\bar F_{Y^2X}/a^2}{3\sqrt{\left(-\bar F_X +2\bar F_X^2\right)\left(\bar F_{Y^2}/2a^2-\bar F_Z/a^2\right)}}\;,\label{lambda3} \\
  \lambda_4 &=& \frac{8\bar F_{X^2Z}/a^2}{3\sqrt{\left(-\bar F_X +2\bar F_X^2\right)\left(\bar F_{Y^2}/2a^2-\bar F_Z/a^2\right)}}\;, \label{lambda4}
 \end{eqnarray}
 while $\lambda_1$ and $\lambda_2$ are defined in \eqref{lambda1}, \eqref{lambda2}.

 Then, the integral that we need to compute (see  Fig \ref{fig:bispectrum}) is 
  \begin{equation}
  \langle \hat\pi_{\vec{k}_1}\hat\pi_{\vec{k}_2}\hat\pi_{\vec{k}_3}\rangle =
  -\int_{\tau_{\rm{min}}}^0\dif\tau_1\int_{\tau_{\rm{min}}}^{\tau_1}\dif\tau_2
  \langle 0 | \left[ \mathcal{H}^{(3)}(\tau_1), \left[ \mathcal{H}^{(2)}(\tau_2),\hat\pi_{k_1}(\tau)\hat\pi_{k_2}(\tau)\hat\pi_{k_3}(\tau)\right]\right] | 0 \rangle \;.
 \end{equation}
 where the third-order interaction $\mathcal{H}^{(3)}$ can be extracted from
 \eqref{operators_bispectrum}.
Let us make some example of the kind of integrals one has to study. 
Consider taking ${\cal H}^{(2)}$ as eq. \eqref{Hint1} and ${\cal H}^{(3)}$ with only the operator $\de_j\sigma_i\de^i\pi\de^j\pi$
 from eq. \eqref{operators_bispectrum}. Then the form of the integral is:
 \begin{eqnarray}
  \langle \hat\pi^3\rangle = 2\,\mbox{Re}\,\Bigg\{&& \!\!\!\!\!\!\!\! \alpha^2\frac{-\lambda_1\lambda_2}{\sqrt{2\left(-\bar F_X + 2\bar F_{X^2}\right)}}\,
			   \int_{-\infty}^0\frac{\dif\tau_1}{(-H\tau_1)^2}\int_{-\infty}^{\tau_1}\frac{\dif\tau_2 }{(-H\tau_2)^3}\,\delta^3(\vec{k}_1+\vec{k}_2+\vec{k}_3) \nonumber \\
		       & & (\vec{k}_1\cdot\vec{k}_2)(\vec{k}_1\cdot\vec{k}_3)\left[ \lambda_1\hat\sigma_{k_1}\hat\sigma^*_{k_1}
		       \hat\pi'_{k_1}\hat\pi^*_{k_1} \left(\hat\pi_{k_2}\hat\pi^*_{k_2}\hat\pi_{k_3}\hat\pi^*_{k_3} - \mbox{c.c}\right)\right. \nonumber \\
		       & & \left.+ \lambda_2 \hat\sigma'_{k_1}\hat\sigma^*_{k_1}
		       \hat\pi_{k_1}\hat\pi^*_{k_1} \left(\hat\pi_{k_2}\hat\pi^*_{k_2}\hat\pi_{k_3}\hat\pi^*_{k_3} - \mbox{c.c}\right)\right] \Bigg\}\;.
 \end{eqnarray}
 
 All the other integrals we compute have a similar structure.
 To keep computations simple and analytical, we assume that the sound speeds are equal, $c_\pi = c_\sigma$.
 We recall that, when computing 
 the power spectrum,
 we were finding log-enhanced contributions.
 We can expect the same amplification effects to occur here 
for the case of a squeezed  bispectrum when a long wavelength mode ($k\rightarrow 0$) is already outside the horizon.
 Then we can consider one of the momenta $k_1 = k_L \to 0$, while the other two are assumed with equal lenght $k_2 \sim k_3 = k_S$, and evaluate the integral from the time when the mode exits the horizon $\tau = - 1/c_\pi k_1$ until the end of inflation.
Summing all the terms arising from the operators \eqref{operators_bispectrum} and focusing only on the leading contributions we obtain
 \begin{equation}
  \langle\hat{\pi}_{\vec{k}_1}\hat{\pi}_{\vec{k}_2}\hat{\pi}_{\vec{k}_3}\rangle = 
  \frac{\alpha^2 H^5}{16 c_\pi^{10} k_L^3k_S^3}\frac{1}{\sqrt{2\left(-\bar F_X+2\bar F_{X^2}\right)}}\,
  \Bigg[9 c_\pi^2 \lambda_1 \lambda_4 + (3\lambda_1+\lambda_2)(\lambda_2+\lambda_3)c_\pi^2 \hat{S}_2 -27 \lambda_1\lambda_2 \hat{S}_1\Bigg]\,
  \log\left(\frac{k_L}{k_S}\right) \;,
 \end{equation}
 where $\hat{S}_1$, $\hat{S}_2$ refer to scalar products between versors of momenta:
 \begin{eqnarray}
  \hat{S}_1 &=& (\hat{k}_1\cdot\hat{k}_2)(\hat{k}_1\cdot\hat{k}_3)+(\hat{k}_2\cdot\hat{k}_3)(\hat{k}_1\cdot\hat{k}_2)+(\hat{k}_1\cdot\hat{k}_3)(\hat{k}_2\cdot\hat{k}_3) \;, \\
  \hat{S}_2 &=& \hat{k}_1\cdot\hat{k}_2 +\hat{k}_2\cdot\hat{k}_3 +\hat{k}_1\cdot\hat{k}_3 \;.
 \end{eqnarray}
 In the squeezed limit, they reduce to:
 \begin{equation}
  \hat{S}_1 = - \cos^2 \theta \;,\qquad \hat{S}_2 = 1-2\cos^2{\theta}\;,
 \end{equation}
 where $\theta$ is the angle between the long and the short wavelengths modes. From
 the three-point functions for $\pi$, as discussed above,
 we can extract the three-point function for the curvature perturbation ${\cal R}$. 
  Using the normalization \eqref{canonical_pi} together with the Friedmann eq. \eqref{friedmann2} and the definition of the speed of sound \eqref{cpi}, we can  write the squeezed limit of the  three-point function for curvature pertubation, up to second order in $\alpha$,  as:
 \begin{eqnarray}
&&  \langle\mathcal{R}^3\rangle_{k_L\to0}  \simeq (2\pi)^{3}\delta^{3}(\vec{k}_{1}+\vec{k}_{2}+\vec{k}_{3})
\alpha^2\frac{\mathcal{P}(k_L)\mathcal{P}(k_S)}{k_{L}^3k_{S}^3}\,\frac{\pi^4}{c_\pi^4}\times\nonumber\\
& &\times \Bigg\{\left[ 9c_\pi^2\lambda_1\lambda_4+ (3\lambda_1+\lambda_2)(\lambda_2+\lambda_3)c_\pi^2 \right]
 +\left[
 27\lambda_1\lambda_2- 2(3\lambda_1+\lambda_2)(\lambda_2+\lambda_3)c_\pi^2\right]\,\cos^2\theta\Bigg\}\log\left(\frac{k_L}{k_S}\right)
 \;,\nonumber\\ 
 \end{eqnarray}
 where we only write the log-enhanced contributions to this quantity. 
 Let us comment on the physical consequences of this result:
 \begin{itemize}
 \item Even if, in the squeezed limit, the curvature three-point function is suppressed by a factor of $\alpha^2$ (a parameter that we consider small) it is nevertheless enhanced by 
 a factor  $\log\left({k_L}/{k_S}\right)$, a quantity that can be of order of the number of e-folds of inflation:
 \be
 \log\left(\frac{k_L}{k_S}\right)
\,\simeq\,N_k\,.
 \ee
 This means that,
 as for the case of the power spectrum, we find a log-enhanced contribution. The same considerations of  Section \ref{sec-pssf} hold here: since we have non-adiabatic fluctuations
   only, these effects are physical and cannot be gauged away with a redefinition of coordinates. Notice that, moreover, the three-point function is enhanced by a large power of the
   sound speed ($1/c_\pi^4$) that can also 
 considerably increase its size, in the case that $c_\pi\,<\,1$.
 \item Interestingly, we find a non-trivial angular dependence of the squeezed limit of the bispectrum.  The squeezed bispectrum can
  be expressed as a sum of two contributions, a monopole plus a quadrupole, with tunable coefficients depending on the parameters $\lambda_i$.
   An angular dependent squeezed 
   bispectrum has been also found in other works in the literature, as Solid Inflation \cite{Endlich:2012pz}, or inflation with vector fields 
   \cite{Shiraishi:2013vja,Bartolo:2012sd,Shiraishi:2012rm,Shiraishi:2012sn,Bartolo:2014hwa,Bartolo:2015dga}, or in models with higher spin fields
    \cite{Arkani-Hamed:2015bza}. In those realizations,
   the coefficients in front of each contributions (monopole and quadrupole) are fixed by the model. 
     In our set-up based  on an EFT approach to inflation, we have been able to identify classes of  operators that allow to obtain more general squeezed limits
     for the bispectrum, with arbitrary coefficients in front of each angular-dependent contribution.   We can then identify a possible origin of 
     these effects    as due to  particular patterns of  space-time diffeomorphism breaking.
     It would be interesting to find concrete models 
      that obtain our operators from a fundamental set-up.

 \end{itemize}

\subsection{Tensor-scalar-scalar bispectra and consistency relations}

In this subsection we examine how breaking space-time diffeomorphisms affects the bispectra
involving tensor and scalar fluctuations. Observables associated with  three-point functions involving tensor modes
are becoming particularly interesting, since they are sensitive to the behavior of gravity at the high scales of inflation, 
  and since the future promises 
  advances  
 in observational
  efforts to detect  primordial tensor modes. 
 We start with a brief review on the present theoretical and observational status 
  of our knowledge of tensor-scalar-scalar bispectra; then we pass to discuss new results we obtain within the EFT of inflation
  with broken space-time diffeomorphisms. 
   
\subsubsection{Motivations}

In light of the amount of precise measurements that are  becoming available,  it is important to
 select the best observables that will 
 clarify the physics responsible for driving  inflation.  Among the  predictions of inflation there is one that affects both the CMB and the LSS of the universe: it is the correlation between primordial scalar and tensor perturbations (gravitational waves) \cite{Maldacena:2002vr}.
  
This tensor-scalar-scalar (TSS) correlation, that is present in all the inflationary models, generates a local power quadrupole in the power spectrum of the scalar perturbations when the wavelength of the tensor mode is much bigger than the scalar one (squeezed limit),  giving rise to an apparent local departure from statistical isotropy. This observable is a useful quantity to discriminate among the plethora of inflationary models. Moreover this long wavelength tensor mode leaves a precise imprints (dubbed \textit{fossils}) on the observed mass distribution of the universe.  
The properties of the correlation functions are dictated by symmetries
  that  have a crucial role in constraining the form of correlation functions, and the corresponding
    consistency relations and their violation \cite{Creminelli:2004yq, Creminelli:2012qr, Goldberger:2013rsa, Berezhiani:2013ewa, Sreenath:2014nka, Kehagias:2012pd, Kehagias:2013xga, Biagetti:2013qqa}. 

In the next years we will see an increased dedicated effort  in trying to detect gravitational waves \cite{Ade:2014xna, Ade:2014gua, Ade:2014afa, Ade:2015tva, AmaroSeoane:2012km, Corbin:2005ny, Kawamura:2011zz}.

In \cite{Dai:2013kra} it has been shown that, in the case of single-clock models, that are space-diffeomorphism invariant, a quadrupole contribution to the TSS is cancelled. In particular, a quadrupole contribution  arises, proportional to  the number of efolds, that is   exactly
 compensated 
 by late-time projection effects that leave a negligible amplitude for the power quadrupole. However, when the conditions of single-clock \cite{Chen:2009zp}, invariance under space diffeomophism \cite{Endlich:2012pz, Kang:2015uha, Emami:2015uva}, slow-roll evolution \cite{Chen:2013aj, Chen:2013eea} are evaded,  then the consistency relation is violated,  the cancellation is not perfect and we get a possibly detectable amplitude for the local power quadrupole. In \cite{Dimastrogiovanni:2014ina} it has been shown how the violations of the slow-roll dynamics in non-attractor inflation and of space-diffeomorphism invariance in Solid Inflation bring to the violation of the consistency relation in the TSS correlation function with a consequent enhancement in the local quadrupole. In the case of non-attractor inflation, the limits from CMB on the statistical isotropy \cite{Ade:2015hxq, Kim:2013gka} constrain the effects on non-observable scales since the transition from the non-attractor phase to the attractor one is found to happen before the time when the current observable universe left the  horizon during the inflationary phase. In the Solid Inflation model, instead, the violation of the consistency relation is related to the violation of the diffeomophism invariance and, more interestingly, the observable anisotropic effects are spread on much smaller scales and so potentially detectable in the next future galaxies surveys. In a recent paper \cite{Dimastrogiovanni:2015pla} the effect of the violation of the consistency relation has been computed in the Quasi-Single-Field model: a two fields model where one of the two has a mass near the Hubble scale $H$. From the non-trivial four-point function they estimate the size of the galaxy survey necessary to detect the effect of the tensor-scalar-scalar consistency violation.

\subsubsection{New results using the EFTI for broken space-time diffeomorphisms}

Our model, violating the invariance under space diffeomorphism, leads to a violations 
  of the consistency relation of the tensor-scalar-scalar correlator, as we are going to discuss.
  
Following \cite{Berezhiani:2013ewa} the three-point function, in the case of a tensor-scalar-scalar interaction, can be re-defined as
\begin{equation}
\left \langle \gamma_{\vec{k}_{1}}^{s} \mathcal{R}_{\vec{k}_{2}} \mathcal{R}_{\vec{k}_{3}}\right\rangle \equiv (2\pi)^3 \delta^{3}(\vec{k}_{1}+\vec{k}_{2}+\vec{k}_{3}) \left \langle \gamma_{\vec{k}_{1}}^{s} \mathcal{R}_{\vec{k}_{2}} \mathcal{R}_{\vec{k}_{3}}\right\rangle' \;,
\end{equation}
where the primed correlator is related to the bispectrum by
\begin{equation}
\left \langle \gamma_{\vec{k}_{1}}^{s} \mathcal{R}_{\vec{k}_{2}} \mathcal{R}_{\vec{k}_{3}}\right\rangle' \equiv \epsilon_{ij}^{s}\;\hat{k}_{2}^i \;\hat{k}_{3}^j \;\mathcal{B}(k_{1},k_{2},k_{3})\;,
\end{equation}
and $\epsilon_{ij}^s$ is the polarization tensor of the tensor mode.

Considering the limit in which the momentum of the tensor $(k_{1})$ is identified with $k_{L}$ (long wavelength) and the momenta of the scalars $(k_{2}, k_{3})$ are identified with $k_{S}$ (short wavelengths), when the consistency relation for the tensor-scalar-scalar correlator is satisfied, the bispectrum can be expressed like 
\begin{equation}
\mathcal{B}(k_{L},k_{S},k_{S})\equiv-\frac{1}{2}P_{\gamma}(k_{L})P_{\mathcal{R}}(k_{S})\frac{\partial \ln P_{\mathcal{R}}(k_{S})}{\partial \ln k_{S}}\;,
\end{equation}
that, in single-field slow-roll models, translates in a quantity proportional to $(n_{s}-4)$, where $n_{s}$ is the scalar spectral index,
as calculated  by Maldacena in \cite{Maldacena:2002vr}. 

In our case from the third order action \eqref{bisFY} we can read that the tensor-scalar-scalar bispectrum has two contributions, that we can schematically write as
\begin{equation}\label{split-bis}
 \mathcal{B}(k_{1},k_{2},k_{3})= \mathcal{B}_{[\gamma \partial \pi \partial \pi]}(k_{1},k_{2},k_{3})+ \mathcal{B}_{[\gamma \partial \dot{\sigma}_{L} \partial \pi]}(k_{1},k_{2},k_{3})\;.
\end{equation}
These two contributions are associated with our novel
 operators corresponding to the fourth and third terms in  \eqref{bisFY}. 
 They add to the other contributions already present in EFTI and Solid Inflation (that we do not consider here), and can be computed using the in-in formalism.  The (normalized) scalar Fourier wavefunctions are defined in \eqref{mode} while for the tensor perturbations we use
\begin{equation} \label{h_mode}
 \gamma_{ij,\vec{k}}=\sum_{s=\pm}\epsilon_{ij}^{s}(\hat{k})\left[c_{\vec{k}}^{s}\;\gamma_{k}+(c^{s})_{-\vec{k}}^{\dagger}\;\gamma_{k}^*\right]\;,
\end{equation}
where ``\textit{s}'' represents the two polarizations of the tensor and the creation and annihilation operators respect the following commutation relation
\begin{equation}
  \left[c_{\vec{k}} ,c^\dagger _{-\vec{k}'}\right]=\left(2\pi\right)^{3}\delta^{(3)}(\vec{k}+\vec{k}')\,\delta_{ss'}\;.
\end{equation}
The scalar wave functions for the two scalar goldstones are given by \eqref{pi_wave} and \eqref{sigma_wave} while for the tensor
we can take the standard expression \eqref{gamma_wave}, since $\alpha^2$-correction to the wave function would be subleading
when considered in this interactions\footnote{The effects of modified wavefunctions
could be interesting, in principle, when considered in the other bispectra which are not proportional to the small $\alpha^2$,
like e.g. the standard $1/c_s^2$ bispectrum.}, that are already proportional to $\alpha^2$.

The effect of the long wavelenght tensor mode on the two scalars is encoded in the squezeed limit ($k_{L}\ll k_{S}$) of the bispectrum $\langle \gamma_{k_{L}}  \pi_{k_{S}} \pi_{k_{S}}\rangle$.
The first contribution that we obtain can be computed at tree-level, following \cite{Maldacena:2002vr}
\begin{equation}
\label{firsttss}
\langle\gamma_{k_{L}}  \hat{\pi}_{k_{S}} \hat{\pi}_{k_{S}}(\tau_{0})\rangle= -\; i \int_{\tau_{min}}^{\tau_{0}}d \tau \left\langle \left[\gamma_{\vec{k}_{L}}(\tau_{0}) \hat{\pi}_{\vec{k}_{S}}(\tau_{0}) \hat{\pi}_{\vec{k}_{S}}(\tau_{0})\;, \mathcal{H}_{\gamma\partial\pi\partial\pi}^{(3)}(\tau)\right]\right\rangle\;,
\end{equation}
and it gives

\begin{figure}%
    \centering
    \subfloat[]{{\includegraphics[width=5cm]{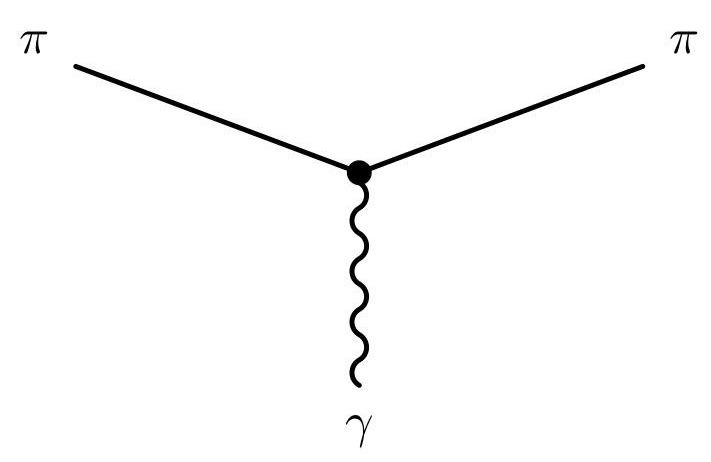} }}%
    \qquad
    \subfloat[]{{\includegraphics[width=5cm]{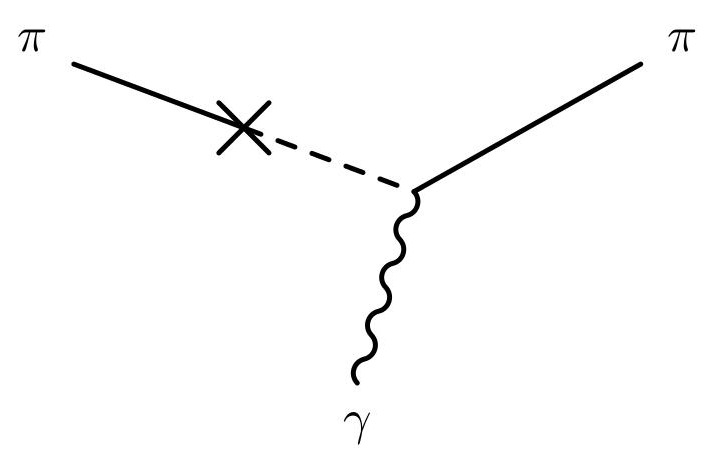} }}%
    \caption{Leading consistency violating contributions to the TSS bispectrum $\langle \gamma \pi\pi\rangle$ }%
    \label{fig:example}%
\end{figure}

\begin{equation}
\mathcal{B}_{[\gamma \partial \mathcal{R} \partial \mathcal{R}]}(k_{L},k_{S},k_{S})= \frac{H^2}{M_{Pl}^2} \frac{\alpha^2}{(F_{X}-2F_{X^2})^2}\frac{F_{Y^2}}{a^2}     \frac{3H^4}{4 c_{\pi}^5} 	\left (\frac{1}{k_{S}^3\,k_{L}^3}\right)\;.
\end{equation}

Rewriting this contribution in terms of the curvature 
and tensor
  power spectra,  we find  a violation of the consistency relation in the tensor-scalar-scalar bispectrum. 
   This since our result is proportional to the quantity 
  $F_{Y}^{2}$ 
   that is  {\it not related} to the scalar spectral tilt; moreover it is not associated with the other observables met so far, so we do not have bounds on its size, although
   for naturalness reasons we do not expect it to be large.
Let us also emphasize that  such contribution to the TSS bispectrum  is 
  a distinctive feature of our set-up that simultaneously breaks time and space diffeomorphisms. 
  
  Interestingly, this is {\it not} the 
    dominant contribution to the bispectrum: 
      from the second interaction term in eq \eqref{split-bis} we have
\begin{equation}
  \langle \gamma_{k_L}\hat{\pi}_{k_2}\hat{\pi}_{k_3}\rangle =
  -\int_{\tau_{min}}^0\dif\tau_1\int_{\tau_{min}}^{\tau_1}\dif\tau_2
 \left \langle  \left[ \mathcal{H}^{(3)}(\tau_1), \left[ \mathcal{H}^{(2)}(\tau_2),\gamma_{k_L}(\tau)\hat{\pi}_{k_2}(\tau)\hat{\pi}_{k_3}(\tau)\right]\right] \right \rangle \;,
 \end{equation}
where $\mathcal{H}^{(2)}$ is given by \eqref{secondint} and
\begin{equation}
  \mathcal{H}_{[\gamma \partial \dot{\hat{\sigma}} \partial \hat{\pi}]}^{(3)}  = - \frac{2\alpha}{M_{Pl}} \lambda_{2}  \int\dif^3x\,a^3 \,\frac{\gamma_{ij}\de^i\hat{\sigma}'\de^j\hat{\pi}}{\sqrt{-\nabla^2}} \;.
\end{equation}
Performing the integral, considering the limit in which the tensor momentum $k_L\to0$ is much smaller than the scalar
momenta $k_S$, the result reads:  
\begin{equation}
\mathcal{B}_{[\gamma \partial \dot{\sigma} \partial \pi]}(k_{L},k_{S},k_{S})=\frac{\alpha^2\lambda_{2} \lambda_{1}}{8M_{Pl}} \frac{H^4}{c_{\pi}^5} 	\left (\frac{1}{k_{S}^3\,k_{L}^3}\right)\log\left(\frac{2c_{\pi}k_{S}}{k_{L}}\right)\,.
\end{equation}
This contribution is the dominant violating contribution to the three-point tensor-scalar-scalar correlation function. Indeed, although it is suppressed by a small parameter
$\alpha$, it 
 has a log-enhancement of the same
  kind we studied in the previous sections, that can be of the order of the number of e-folds.
   
Going back to the original correlator $\langle \gamma  \pi \pi\rangle $, and considering only the leading contribution, rewriting it in terms of the curvature perturbation we find 
\begin{equation}\label{tss2}
\mathcal{B}_{[\gamma \partial \mathcal{R} \partial \mathcal{R}]}(k_{L},k_{S},k_{S}) \supset \frac{\alpha^2}{ M_{Pl}^2}\frac{\lambda_{2} \lambda_{1}}{\left(-F_X+2F_{X^2}\right)} \frac{H^6}{4 c_{\pi}^5} 	\left (\frac{1}{k_{S}^3\,k_{L}^3}\right)\log\left(\frac{2c_{\pi}k_{S}}{k_{L}}\right)\,.
\end{equation}
or in terms of the scalar and tensor power spectra

\begin{equation}
 \mathcal{B}_{[\gamma \partial \mathcal{R} \partial \mathcal{R}]}(k_{L},k_{S},k_{S}) \supset P_{\gamma}(k_{L})P_{\mathcal{R}}(k_{S})  \frac{\alpha^2\lambda_1\lambda_2}{4 c_{\pi}^2}   \log\left(\frac{2c_{\pi}k_{S}}{k_{L}}\right) \;,
 \end{equation}
and this our final result.

As anticipated, when a long wavelength tensor mode correlates with the density (scalar) fluctuations in the tensor-scalar-scalar squeezed bispectrum a local power quadrupole is generated. This contribution, that appears like a departure from statistical isotropy, shows an infrared-divergent behaviour, that becomes negligible $\mathcal{O}\left(\frac{{k_{L}}^{2}}{{k_{S}}^{2}}\right)$ when late time projection effects are taken into account \cite{Dai:2013kra} in the case when the consistency relation is satisfied, but {\it not} in our case.

The local quadrupole $Q$ enters in the power spectrum as anisotropic contribution 
\begin{equation}
\label{quad}
P_{\mathcal{R}}(\vec{k}_{S})|_{\gamma(\vec{k}_{L})}=P_{\mathcal{R}}(\vec{k}_{S})\left[1+\mathcal{Q}_{ij}^{p}(\vec{k}_{L})\,{\hat{k}}_{S}^{i}\,{\hat{k}}_{S}^{j}\right]\;,
\end{equation}
and it is defined as the ratio between the consistency-relation-violating contribution of the tensor-scalar-scalar bispectrum  $\mathcal{B}_{\rm{cv}}(k_{L},k_{S},k_{S})$ and the power spectra of the scalar and tensor modes. Estimating the variance of the quadrupole, that is the observable quantity, it is possibile to extract informations about the parameters of the theory. In our case its value is not so informative in putting constraints on the model with respect to the previous observables.\\
On the other hand a long wavelength tensor mode can leave ``\textit{fossil}'' imprints also on the Large Scale Structure. In this case a tensor mode with wavelength smaller than our observable universe is considered and from an estimator for the tensor power spectrum and its variance it is possibile to extract informations on the minimum size of the galaxy survey on which the tensor can be detected. We report an estimate of the survey size in Appendix \ref{app-C} deserving a careful parameter space analysis of the theory in future work. From the estimate we can see that in the next galaxy survey, like EUCLID or even better in $21$-$\rm{cm}$ will be possibile to put bounds and test our theory. So we want to emphasize that even though some (null) searches for power asymmetry in the CMB \cite{Ade:2015hxq, Groeneboom:2008fz} and Large Scale Structure \cite{Pullen:2010zy} have already been done, much effort is needed because we have seen how this signatures becomes important in order to rule out inflationary models and also to give informations on the pattern of symmetries in the early universe.

\section{Conclusions}\label{sec-concl}

Using an effective field theory approach to inflation (EFTI),  we discussed the case for breaking both time and space
diffeomorphism   invariance during inflation. While
 EFTI has been largely  
 applied in the past to study inflationary set-ups with broken time reparameterisation invariance,
 we pointed out new  effects occurring in a set-up where   both space and time diffeomorphisms are simultaneously broken during inflation.  
We provided physical 
 motivations for considering such system,
  explained how to build an EFT to analyse it. 
 We showed  that it can lead to new interesting observational consequences, as a blue spectrum of gravitational
waves, 
 enhanced 
amplitudes  for bispectra, and moreover
  a characteristic angular dependence between the wavevectors in
  the
  squeezed limits 
 of bispectra.
Although we break space-time diffeomorphisms acting on the coordinates, for simplicity we focus on scenarios that 
 preserve internal rotational and translational symmetries in the space of fields
 driving inflation.
 We built  actions describing the physics 
 of Goldstone bosons  associated with our pattern of symmetry breaking, and   the dynamics of vector and tensor modes.
  In our scenario  we find two 
  scalar Goldstone bosons: one scalar $\pi$
  associated with the breaking of time reparameterisation, and one scalar $\sigma$ -- playing the role of a phonon -- associated
  with the  breaking of space translations. Tensor fluctuations enjoy new types of interactions with themselves and
  with other sectors, associated with operators allowed by the fact that we fully break diffeomorphism invariance.

We discussed observables relative  both to  scalar   and tensor spectra, associated with two and three
point functions among fluctuations. 
  We studied both power spectra and bispectra. We determined two   broad physical effects that are distinctive  of our set-up:

 \begin{itemize}
 \item
  The first  is specific of the scalar sector, and exploits the new couplings between the two scalar Goldstone
 modes of broken symmetries. We found potentially  large contributions to inflationary observables,
   that can give sizeable effects even in the limit of small breaking of space diffeomorphisms.
 Such contributions lead to a change in the  amplitude of the power spectrum of scalar fluctuations, and, more interestingly,
 direction dependent contributions (of the form parametrized in~\cite{Shiraishi:2013vja}) to the squeezed limit of the scalar and tensor bispectra. We discussed the physical
  consequences of 
  these findings, pointing our similarities and differences with previous results in the literature.
  \item The second effect is instead more specific of the tensor sector, and exploits novel possibilities
  for   tensors to couple with themselves and with  scalars. Such possibilities  are associated with  operators that
  are allowed only if we break also  space reparameterisation invariance during inflation.  They can lead to a blue
  spectrum for gravitational waves, and
 moreover to 
 a  particular structure for the squeezed limit of tensor-scalar-scalar bispectra, that violate single field consistency
 relations and can lead to distinctive observable signatures.
 \end{itemize}
 
 \bigskip 
 Our findings show that the effective field theory for inflation is very well suited for studying situations where all diffeomorphisms are broken during inflation,
 and suggest new distinctive features for inflationary observables.
 Future work on this subject can proceed  in two different directions. At the observational level, more work is needed to fully characterize the properties of $n$-point functions in these scenarios -- possibly not only three but also higher point functions. It would be interesting to find distinctive  consistency 
  relations associated with our symmetry breaking pattern, or new observables that specifically test particular
  features of breaking the space diffeomorphism. At the theoretical level, it would be interesting to find 
   new  examples of inflationary models that break all space-time diffeomorphisms (that presumably generalize
   Solid Inflation) and that then can concretely realize the new observable consequences that we pointed out using a
   model independent EFT approach. This amounts to characterize physical
    scenarios -- as scalars or higher spin fields --  that can break space diffeomorphisms during inflation. 
    Finding explicitly realizations of such set-ups would help also to 
    understand what happens {\it after inflation} to whatever physics  is responsible for breaking all diffeomorphisms.  And possibly, find a mechanism to  recover space diffeomorphism invariance, and  fully  turn off the dynamics
    of the second Goldstone boson $\sigma$  after inflation ends.

\smallskip

\subsection*{Acknowledgments}

It is a pleasure to thank Matteo Fasiello and Marco Peloso for useful discussions and feedbacks.  
The work of N.B. was supported in part by ASI/INAF Agreement I/072/09/0 for the Planck LFI Activity of Phase E2.
 G.T. was supported by 
 an STFC Advanced Fellowship ST/H005498/1.
 
\bigskip

\appendix
 
 \section{Mixing with gravity and decoupling Limit} \label{app-A}
 
 In this section we will show why taking the decoupling limit is a consistent approximation in the case under study.
 Similarly to the equivalence theorem for massive gauge bosons, we expect that the physics of the Goldstone decouples from the
 transverse modes above a certain energy scale, $E_{mix}$. For example, in a non-Abelian gauge theory,
 \begin{equation}
  \mathcal{L}=-\frac{1}{4}F_{\mu\nu}^2-\frac{1}{2}(\de_\mu\pi)^2-\frac{1}{2}m^2A_\mu^2+i m\de_\mu\pi A^\mu\;,
 \end{equation}
 where $m^2=f_\pi^2g^2$, the decoupling limit is reached taking the limit $g\to0$, $m\to0$ with $f_\pi$. Therefore,
 for energies $E>m$, the mixing between the Goldstone and the gauge modes becomes irrelevant and the two sectors effectively
 decouple.
 
 Just like the gauge theory analogy, in our case we can find a decoupling limit which corresponds to the limits
 $\mpl\to\infty$, $\dot H\to0$ with $\mpl^2\dot H$ fixed\footnote{This is the same as the previous example with the
 indentifications $g\to\mpl^{-1}$ and $m\to\dot H$.}. To see that taking this limit in our case effectively lead to the decoupling
 of Goldstones and gravity, let us consider first a simplified case, where all the derivatives of $F$ are zero but $F_X$.
 When expanding the the operator $X$ \eqref{intro:blocks} according to \eqref{goldstoneexpansion}, one obtains
 \begin{equation}
  X=(1+\dot\pi)^2g^{00}+2\de_i\pi g^{0i} + \de_i\pi\de_j\pi g^{ij} \;.
 \end{equation}
 Substituted back into the action \eqref{intro:action}, the leading mixing of the Goldstone $\pi$ with gravity will be of the
 form:
 \begin{equation}
  \bar F_X \dot\pi \delta g^{00} \;.
 \end{equation}
 After canonical normalization $\pi_c \sim\sqrt{-\bar F_X}\pi$, $\hat g^{00}_c \sim \mpl \delta g^{00}$ (which gives to the fields
 the dimension of a mass), we can see that taking the decoupling limit $\mpl\to\infty$, $\dot H\to0$ with $\mpl^2\dot H$ implies
 that mixing terms becomes irrelevant with respect to the standard kinetic term $\pi_c^2$ and can be neglected above a certain energy
 $E_{mix}$:
 \begin{equation}
  E_{mix}^2\sim \frac{\bar F_X}{\mpl^2} \sim \frac{\bar F_X}{\mpl^2H^2}H^2\sim\epsilon H^2\ll H^2\;,
 \end{equation}
 where we have used \eqref{friedmann1} and \eqref{epsilon}.
 Therefore as long as $E_{mix}$ is smaller than $H$, we can safely neglect mixing terms, as they would appear in the
 action suppressed by powers of $(E_{mix}/H)^2\sim\epsilon$, since $H$ is our infrared cutoff.
 The same will happen for the other terms present in the action, but in general the answer depends
 on which operators are present and significant. For example, from $Z^{ij}$, after the canonical normalization
 $\sigma_c\sim\sqrt{-\bar F_Z+\bar F_{Y^2}/2}$, one has:
 \begin{equation}
  \bar F_Z \dot\sigma^ig^{0j}\delta_{ij}\qquad\Longrightarrow\qquad
  E_{mix} \propto \left\{ \begin{array}{lcl}
		      \displaystyle\frac{\alpha \sqrt{-\bar F_Z/a^2}}{\mpl}\;, & \mbox{for} & |\bar F_{Y^2}| \lesssim |\bar F_Z| \\
		      \displaystyle\frac{- \alpha \bar F_Z/a^2}{\mpl\sqrt{\bar F_{Y^2}/a^2}}\;, & \mbox{for} & |\bar F_Z| \ll |\bar F_{Y^2}| \\
                      \end{array} \right.
 \end{equation}
 In the first case, as $\alpha\lesssim1$,
 \begin{equation}
  \frac{E_{mix}^2}{H^2} \sim \frac{-\alpha^2\bar F_Z/a^2}{\mpl^2H^2} \lesssim \epsilon \;,
 \end{equation}
 where we have used \eqref{epsilon}.
 In the second case, $\bar F_Z \ll \bar F_{Y^2}$, one can find a similar expression too:
 \begin{equation}
  \frac{E_{mix}^2}{H^2} \sim \frac{-\alpha^2\bar F_Z/a^2}{\mpl^2H^2}\frac{\bar F_Z}{\bar F_{Y^2}} \lesssim \frac{-\alpha^2\bar F_Z/a^2}{\mpl^2H^2} \lesssim \epsilon \;.
 \end{equation}
 Also in this case $E_{mix}$ is smaller than $H$ and the decopling limit can be safely taken.
 However, if for example one has $|\bar F_{Y^2}|\gg |\bar F_Z|$, then, looking at the expansion of the operator $Y_iY^i$ one can
 see that working in the decoupling limit can restrict the range of the allowed parameters:
 \begin{equation}
  \bar F_{Y^2} \dot\sigma^ig^{0j}\delta_{ij} \qquad\Longrightarrow\qquad E_{mix}^2\sim\frac{\alpha^2\bar F_{Y^2}/a^2}{\mpl^2}\;,
 \end{equation}
 which is lower than $H$ only if $\bar Z\bar F_{Y^2}/\mpl^2H^2\ll1$.

 \section{Strong coupling}\label{app-B}
 
 As it is usual in effective field theories, the non-renormalizable self-interactions of the Goldstone fields will become strongly
 coupled at a certain energy scale, $\Lambda_{st}$, beyond which the theory ceases to make sense and new physics must enter.
 In our case, we have to make sure that $\Lambda_{st}\gg H$ so that the theory is weakly-coupled in the energy regime we are
 interested in.
 
 Stronger interactions are related to smaller kinetic energy: indeed if the time-kinetic terms in \eqref{scalaraction} have
 prefactors of order $\sim \epsilon$, we would canonically normalize the fields (collectively denoted with $\pi$ for simplicity)
 like $\epsilon F (\de\pi)^2 \sim (\de{\hat{\pi}})^2$ and
 inverse power of $\epsilon$ will appear in higher order terms, which would mean stronger interactions or,
 equivalently, a lower strong coupling scale. Of course, if the coefficients of the kinetic terms were bigger or the
 coefficients of higher-oder terms were smaller, interactions would be accordingly weaker. As we have to impose a lower bound on
 $\Lambda_{st}$, from now on we will focus
 only on the ``worst possible case'', when prefactors of time-kinetic terms are as small as $\sim \epsilon F$, while interactions,
 which are proportional to higher derivatives of $F$ with respect to the operators $X$, $Y^i$ and $Z^{ij}$,
 are as big as $F$ itself.
 
 Let us first consider the case with speeds of sound very close to unity. In this case, after canonical normalization,
 we can directly read the strong coupling scale as the scale suppressing higher-order operators
 in the action, $(\partial {\hat{\pi}})^3/\Lambda^2$. The result is simply:
 \begin{equation}
  \Lambda^4_{st} \simeq \epsilon^3 F \;.
 \end{equation}
 If the speed of sound are non-relativistic, the cut-off can not be immediately read from the action as there is an hierarchy
 between time and spatial derivatives and the theory is not Lorentz invariant. Let's assume for simplicity
 that $c_\pi \simeq c_\sigma=c_s\ll1$. We can rescale the time coordinate \cite{Senatore:2010jy, Endlich:2010hf} as $t\to t/c_s$,
 in order to remove this hierarchy. The quadratic action has now the form:
 \begin{equation}
  S_2\simeq\int\dif^4 x\sqrt{-g}\,\epsilon F c_s (\de_\mu \pi)^2 \;,
 \end{equation}
 and the fields would be normalized as $\epsilon c_s F \pi = {\hat{\pi}}$. Schematically, after canonical normalization, the cubic
 interactions will have the form
 \begin{equation}
  S_3\simeq\int\dif^4x\sqrt{-g}\,\frac{(\de {\hat{\pi}})^3}{c_s^{5/2}\epsilon^{3/2}\, F^{1/2}} \;.
 \end{equation}
 where in the denominator the strong coupling \emph{momentum} scale appears. We can obtain the energy scale $\Lambda_{st}$
 multiplying by an extra $c_s$. The result is:
 \begin{equation}
  \Lambda_{st}^4\simeq \epsilon^3c_s^9\,F \;.
 \end{equation}
 As we said, our theory is under control if $\Lambda_{st}\gg H$, which will give the constraint:
 \begin{equation}
  \epsilon\, c_s^3 \gg \left(\frac{H}{\mpl}\right)^{2/3} \;,
 \end{equation}
 where we have used the Friedmann equation \eqref{friedmann1}. This is only an order-of-magnitude estimate and,
 given the many possible combinations of free parameters that are allowed in our action, this constraint can also be
 not very restrictive. However it is still an important bound to respect for the consistency of the theory.

\section{Tensor fossil estimation} \label{app-C}

{
In order to extract precise informations about the size of the galaxy surveys on which the long wavelength tensor mode can leave ``\textit{fossil}'' imprints we need to use the optimal estimator for the tensor power spectrum constructed in \cite{Jeong:2012df}. In this case we consider a tensor mode which wavelength is smaller than the size of the observable universe and then we compute the variance of the optimal estimator
\begin{equation}
\label{var}
\sigma^{-2}_{\gamma}=\frac{1}{2}\sum_{\vec{k}_{L},p}\left[ k_{L}^{3} P_{p}^{n}(k_{L})\right]^{-2} \,\, ,
\end{equation}
where ``p'' are the two polarizations of the tensor, $P_{p}^{n}$ is the noise power spectrum, defined as the ratio between the consistency violating contribution to the bispectrum and the total power spectrum
\begin{equation}\label{noise}
P_{p}^{n}(k_{L})=\left[\sum_{\vec{k}_{S}}\frac{|\mathcal{B}_{
	  {\rm cv}}(k_{L},k_{S},|\vec{k}_{L}-\vec{k}_{S}|)\epsilon_{ij}^{p}\hat{k}_{S}^{i}\hat{k}_{LS}^{j}|^{2}}{2 V P_{\gamma}^{2}(k_{L})P^{tot}(k_{S})P^{tot}(|\vec{k}_{L}-\vec{k}_{S}|)}\right]^{-1} \, \, ,
\end{equation}
where  $V\equiv \frac{(2\pi)^3}{k_{min}^{3}}$ is the total volume of the survey and $\epsilon_{ij}^{p}$ is the polarization tensor.\\
The total power spectrum, that is the measured one, includes both the noise and the signal, $P^{tot}(k)=P(k)+P^{n}(k)$. The bispectrum can be written in terms of a function $f(\vec{k}_{1}, \vec{k}_{2})$, that describes the coupling of the soft mode, and the ``long" mode power spectrum $P(k_{L})$
\begin{equation}
B(\vec{k}_{L},\vec{k}_{1},\vec{k}_{2})=P(k_{L})f(\vec{k}_{1},\vec{k}_{2})\epsilon_{ij}^{p}(\hat{k}_{L})\hat{k}_{1}^{i}\hat{k}_{2}^{j}=\mathcal{B}(k_{L},k_{1},k_{2})\epsilon_{ij}^{p}(\hat{k}_{L})\hat{k}_{1}^{i}\hat{k}_{2}^{j}\,,
\end{equation}
 in such a way that the noise power spectrum becomes \cite{Jeong:2012df}
\begin{equation}
P_{p}^{n}(k_{L})=\left[\sum_{\vec{k}_{S}}\frac{|f(\vec{k}_{S},\vec{k}_{L}-\vec{k}_{S})\epsilon_{ij}^{p}\,k_{S}^{i}\,(k_{L}-k_{S})^{j}|^{2}}{2 V P^{tot}(k_{S})P^{tot}(|\vec{k}_{L}-\vec{k}_{S}|)}\right]^{-1} \, \, .
\end{equation}
$k_{L}$ and $k_{S}$ are the wave number of the long wavelength mode and the short wavelength one.\\
The function $f(\vec{k}_{1}, \vec{k}_{2})$ can be easily read from the tensor-scalar-scalar bispectrum \eqref{tss2}
\begin{equation}
f(k_{S}, k_{L})= \frac{C\; P(k_{S})}{k_{S}^2} \log \left(\frac{2 c_{\pi}k_{S}}{k_{L}}\right)\hskip1cm  , \hskip1cm  C=\frac{\alpha^2}{4 c_{\pi}^2}\lambda_{1}\lambda_{2}\;,
\end{equation}
where we see the novel dependence from the number of modes in the survey. Even if we know that in our case the tensor power spectrum is not exactly scale invariant, at lowest order in $\alpha$ we can assume a nearly scale invariant fiducial power spectrum with amplitude $A_{\gamma}$, $P_{\gamma}=A_{\gamma} k_{L}^{n_{\gamma}-3}$ with $n_{\gamma}\simeq 0$.  \\
Assuming $P^{(0)}(k_{S})/P^{tot}(k_{S})\simeq 1$ ( ``correction" to the power spectrum much smaller than 1) if $k_{S}\le k_{max}$ and equal to zero otherwise, where $k_{max}$ is the largest wavenumber that allows for a large signal-to-noise measurement, we compute the noise power spectrum. Plugging this quantity in \eqref{var} and considering that a signal is detected if it has an amplitude larger than $3\sigma$ we obtain

\begin{equation}
3\sigma_{\gamma}\simeq\frac{18 \sqrt{3}\;\pi^{3/2}}{C^2}\left(\frac{k_{\rm{min}}}{k_{\rm{max}}}\right)^{3}\log \left(\frac{2 c_{\pi}k_{\rm{min}}}{k_{\rm{max}}}\right)^{-2}\;.
\end{equation}
Inverting this relation we can find the size of the galaxy survey necessary to detect at $3\sigma$ the imprints of primordial tensor mode with a given amplitude $A_{\gamma}$. The estimation, as we can see, would be model dependent and require an improved parameter space analysis of the model, but in order to have a rough estimation, if we assume  $c_{\pi}\simeq 10^{-1}$, for the parameter $C$ in the range $(0.1-1)$, one finds that a detectable primordial tensor mode with an amplitude $A_{\gamma}\simeq 2\times 10^{-9}$, that is a value close to the current upper limits,  requires a survey with size in the range $\frac{k_{max}}{k_{min}}\sim \left(4000-1000\right) $, a value that can be achievable with the next survey like $21$-$\rm{cm}$
\cite{Loeb:2003ya}.}

\end{document}